\documentclass[prd,twocolumn,superscriptaddress,floatfix,nofootinbib]{revtex4-2}
\usepackage{amsmath,amssymb,amsfonts,amsthm}
\usepackage{graphicx}
\usepackage{enumitem}
\usepackage{bm}
\usepackage{rotating}
\usepackage{hyperref}
\usepackage{pbox}
\usepackage{array}
\usepackage{mathtools}
\usepackage{leftidx}
\usepackage{lmodern}
\usepackage[normalem]{ulem}
\usepackage{braket}
\usepackage{tensor}
\usepackage[usenames,dvipsnames]{color}
\usepackage{float}
\usepackage{footnote}
\usepackage{setspace}
\usepackage{CJKutf8}

\newcommand{\tcr}[1]{\leavevmode{\color{Black}{#1}}}

\newcommand{\dd}{\text{d}}
\newcommand{\rr}[1]{\left(#1\right)}
\newcommand{\bx}{{\bm{x}}}
\newcommand{\by}{{\bm{y}}}

\newcommand{\pdag}{{\phantom{\dagger}\!}}
\newcommand{\bk}{{\bm{k}}}
\newcommand{\sx}{\mathsf{x}}
\newcommand{\sy}{\mathsf{y}}

\newcommand{\ii}{\text{i}}
\DeclareMathOperator{\tr}{\text{Tr}}

\newcommand{\ab}{{\textsc{ab}}}
\newcommand{\skri}{\mathcal{I}}

\begin{document}

\title{When entanglement harvesting is not really harvesting}

\author{Erickson Tjoa}
\email{e2tjoa@uwaterloo.ca}
\affiliation{Department of Physics and Astronomy, University of Waterloo, Waterloo, Ontario, N2L 3G1, Canada}
\affiliation{Institute for Quantum Computing, University of Waterloo, Waterloo, Ontario, N2L 3G1, Canada}

\author{Eduardo Mart\'in-Mart\'inez}
\email{emartinmartinez@uwaterloo.ca}
\affiliation{Institute for Quantum Computing, University of Waterloo, Waterloo, Ontario, N2L 3G1, Canada}
\affiliation{Department of Applied Mathematics, University of Waterloo, Waterloo, Ontario, N2L 3G1, Canada}
\affiliation{Perimeter Institute for Theoretical Physics, 31 Caroline St N, Waterloo, Ontario, N2L 2Y5, Canada}


\date{\today}

\begin{abstract}

We revisit the entanglement harvesting protocol when two detectors are in causal contact. We study the role of field-mediated communication in generating entanglement between the two detectors interacting with a quantum field. We provide a quantitative estimator of the relative contribution of communication versus genuine entanglement harvesting. For massless scalar fields in flat spacetime, we show that when two detectors can communicate via the field, the detectors do not really harvest entanglement from the field, and instead they get entangled only via the field-mediated communication channel. In other words, in these scenarios the entanglement harvesting protocol is truly ``harvesting entanglement'' from the field only when the detectors are not able to communicate. In contrast, for massive scalar fields both communication and genuine harvesting contribute equally to the bipartite entanglement when the detectors are causally connected. These results emphasize the importance of taking into account the causal relationships between two parties involved in this relativistic quantum information protocol before we can declare that it is truly entanglement harvesting. 

\end{abstract}

\maketitle

\section{Introduction}


It is by now well-known that quantum fields can display quantum correlations between any two disjoint spacetime regions \cite{summers1985bell,summers1987bell,Higuchi2017wedges}. These correlations can be extracted by two localized quantum probes (e.g., Unruh-DeWitt (UDW) detectors \cite{Unruh1979evaporation,DeWitt1979}, atoms coupled to the electromagnetic field~\cite{Lopp2021deloc,pozas2016entanglement}, etc.)---even when they are spacelike separated---via a protocol that has become known as \textit{entanglement harvesting} (see, e.g.,~\cite{Valentini1991nonlocalcorr,reznik2003entanglement,pozas2015harvesting,pozas2016entanglement}). In the context of relativistic quantum information, entanglement harvesting has been used to probe physical phenomena including the underlying spacetime geometry \cite{VerSteeg2009entangling,Ng2018AdS,henderson2019AdSharvesting,Smith2020harvestingGW,Gray2021shockwave}, topology~\cite{smith2016topology}, presence of horizons or boundary conditions \cite{henderson2018harvestingBH,cong2019entanglement,cong2020horizon,Tjoa2020vaidya,Gallock-Yoshimura2021freefall}, indefinite causal order \cite{Henderson2020temporal} or centre of mass delocalization \cite{Nadine2021delocharvesting}. 

When two detectors are spacelike separated, it is clear that the entanglement they acquire has to necessarily come from the harvesting of correlations that exist in the field, since the two detectors cannot communicate. However, it is not uncommon in the study of entanglement harvesting protocols to consider the regimes where the detectors are causally connected~(see, among many others, \cite{olson2012timelikeharvesting,Sanders2016harvesting-position,liu2021entanglement,pozas2015harvesting,pozas2016entanglement,Nadine2021delocharvesting}, etc). However, the question whether these detectors are harvesting correlations between timelike or lightlike separated regions of the field is not obviously clear, since causally connected detectors can potentially get entangled through two mechanisms: (1) genuinely harvesting correlations from the field, or (2) communicating with each other via the field without harvesting any pre-existing field correlations. 

Ideally, we would like to be able to separate the contribution to the entanglement acquired between two detectors into two components: the entanglement that is genuinely harvested, and the entanglement that is generated through the communication of the two detectors via the field. Unfortunately, so far there was no quantitatively clear way to distinguish both contributions. To address this, in this paper we are going to propose a quantitative estimator of the contribution that communication has to the entanglement acquired by detectors in causal contact.

We will do so by noting that the correlations acquired between the detectors can be separated in two (sub-additive) contributions: the contribution coming from the real part of the Wightman function of the field (the expectation of the anti-commutator) and the one coming from its imaginary part (the expectation of the commutator). We will argue that the relative contribution of the commutator part to the total acquired entanglement will give a faithful estimator of how much the entanglement is coming from communication and how much from genuine harvesting.

We will give mainly two reasons for this: on the one hand, it has been shown that the leading order contribution to communication between the detectors is exclusively given by the field commutator, and this contribution enters the final state of the detectors at the same leading order as the harvesting contribution \cite{Jonsson2014cavityQED,Causality2015Eduardo,Casals2020commBH}. Secondly, and more importantly, the commutator contribution is \textit{state-independent}: any entanglement that comes from the commutator contribution cannot be ascribed to the field state (hence unrelated to entanglement structure of the field theory) and will be the same even if there are no  correlations in the field state.  Consequently, genuine vacuum entanglement harvesting must necessarily come from the (state-dependent) anti-commutator contribution. In the cases where the commutator contribution constitutes near the 100\% of the entanglement, one can safely claim that the entanglement acquired through the protocol is not harvested. Again, this is because in that case the entanglement of the detectors would be the same even if we replace the sate of the field by another one with no correlations whatsoever, since the commutator expectation is state-independent. This is particularly important in light of recent results where one can suspect that a significant amount of the detectors' entanglement may be due to field-mediated communication (see, e.g., \cite{Tjoa2020vaidya,Gallock-Yoshimura2021freefall,robbins2020entanglement}). Field-mediated communication is also expected to contribute to the bipartite entanglement when one implements indefinite causal ordering between two detectors' switching times \cite{Henderson2020temporal}, or when the detectors have spatial smearings that make the detectors causally connected (e.g., some regimes of \cite{pozas2015harvesting,Nadine2021delocharvesting}).

In this paper we will proceed as follows: First, we build a quantitative measure that distinguishes harvesting from communication-mediated entanglement. Then, we will show how the behaviour of the field commutator that governs communication between the two detectors depend on spacetime dimensions and the mass of the field. In doing so, we will take into account the decay laws of the real and imaginary parts of the Wightman function, and how they are affected by the spacetime dimension and the strong Huygens principle~\cite{McLenaghan1974huygens,CourantHilbert1989Vol2ch6}. We will see that as a general rule for flat spacetime, when there is causal contact between the detectors there is little or no entanglement harvesting; rather, the entanglement acquired by the detectors will be mostly due to their communication through the field. This is expected to be true {at least} for conformally coupled massless fields in (conformally flat) curved spacetimes. {We will briefly discuss on what can be expected in spacetimes which are not conformally flat such as Schwarzschild geometry \cite{Casals2021commBH2} or spacetimes which admits no conformally flat slicing such as Kerr geometry \cite{Price2000kerrslice,DeFelice2019kerrslice}.}  
Finally, we will also study how the field mass affects these results, and show that the results are the same whether the detectors are compactly supported on spacetime or whether they have Gaussian tails in their switching functions.

This paper is organized as follows. In Section~\ref{sec: UDW} we outline the UDW model and the entanglement harvesting protocol. In Section~\ref{sec: setup} we review the Wightman function, its splitting into anti-commutator and commutator and the strong Huygens' principle. In Section~\ref{sec: comm} we calculate explicitly the density matrix elements for two detectors that interact with a scalar field, and build the communication-mediated entanglement estimator. In Section~\ref{sec: results} we present our main results for massless scalar fields in $(1+1),(2+1)$ and $(3+1)$ dimensions. In Section~\ref{sec: further-results} we discuss how the results change in higher spacetime dimensions and when the field is massive, ending the section with a comparison between the cases of compact switching vs non-compact switching. Throughout this paper we will use natural units $c=\hbar=1$, ``mostly plus'' metric signature and $\sx=(t,\bx)$ is used as a shorthand for spacetime points.

\section{Entanglement harvesting protocol}
\label{sec: UDW}

Two detectors interacting with a quantum field can get entanglement through two mechanisms: they can exchange signals, or they can \textit{swap} the entanglement already present in the state of the quantum field~\cite{summers1985bell,summers1987bell}, allowing them to get entangled even when they are spacelike separated~\cite{reznik2003entanglement,Valentini1991nonlocalcorr,pozas2015harvesting}. Let us summarize the simplest entanglement harvesting protocol.

Let us consider a quantized scalar field of mass $m$ in $(n+1)$-dimensional Minkowski spacetime. In terms of plane-wave modes, we can write the field as
\begin{align}
    \hat\phi(t,\bx) &= \int{\frac{\dd^n\bk}{\sqrt{2(2\pi)^n\omega_\bk}}\rr{\hat a_\bk e^{-\ii\omega_\bk t+\ii\bk\cdot\bx }+\text{H.c.}}}\,,
\end{align}
where $\omega_\bk=\sqrt{|\bk|^2+m^2}$ is the relativistic dispersion relation and the annihilation and creation operators obey the canonical commutation relations $[\hat a^{\pdag}_\bk,\hat a^{\dagger}_{\bk'}]=\delta^n(\bk-\bk')$. Here, the canonical quantization of the field is carried out with respect to inertial observers with coordinates $\sx=(t,\bx)$, where $t$ is the standard Killing time.

Consider two observers Alice and Bob, each carrying a pointlike Unruh-DeWitt detector consisting of a two-level system interacting locally with the quantum field. The total interaction Hamiltonian is given by
\begin{align}
    \hat H^t_I(t) &= \frac{\dd\tau_\textsc{a}}{\dd t}\hat H_{\textsc{a}}^{\tau_\textsc{a}}(\tau_\textsc{a}(t))+ \frac{\dd\tau_\textsc{b}}{\dd t}\hat H_{\textsc{b}}^{\tau_\textsc{b}}(\tau_\textsc{b}(t))\,,\\
    \hat H^{\tau_j}_{j} &= \lambda_j\chi_j(\tau_j)\hat\mu_j(\tau_j)\otimes \hat\phi(t(\tau_j),\bx(\tau_j))\,.
\end{align}
The superscript on $\hat H^t$ means that the Hamiltonian generates time translations with respect to the Killing time $t$ and $\tau_j$ is the proper time of detector $j=\textsc{A}, \textsc{B}$. The switching function $\chi_j(\tau_j)$ prescribes the duration of interactions, and for simplicity we assume that $\chi_j$ is real. Each detector's monopole moment $\hat\mu_j$ given by
\begin{align}
    \hat\mu_j(\tau_j) = \ket{e_j}\!\bra{g_j}e^{\ii\Omega_j\tau_j}+\ket{g_j}\!\bra{e_j}e^{-\ii\Omega_j\tau_j}\,,
\end{align}
where $\{\ket{g_j},\ket{e_j}\}$ are ground and excited states of detector $j$. For simplicity we will consider identical detectors so that $\lambda_j=\lambda$ and $\Omega_j=\Omega$.  

In this work we will consider detector trajectories that are at rest relative to the quantization frame $(t,\bx)$, we can replace $\sx_j(\tau_j)=(t(\tau_j),\bx_j(\tau_j))$ in Eq.~\eqref{eq: Lij} and \eqref{eq: M-nonloc} with $(t_j,\bx_j)$ where $\bx_j$ are constants for $j=A,B$. Since the detectors are taken to be pointlike, without loss of generality we set the trajectories to be 
\begin{align}
    \sx_\textsc{a}(t) = (t,0,0,0)\,,\quad \sx_\textsc{b}(t) = (t,L,0,0)\,,
\end{align}
where $L=|\bx_\textsc{b}-\bx_\textsc{a}|$ is the proper distance between the detectors.

The detector-field interaction for a given initial state $\hat{\rho}_0$ is implemented by unitary time evolution $\hat\rho = \hat U\hat{\rho}_0\hat U^\dagger$, where the time evolution operator $U$ is given by the time-ordered exponential
\begin{align}
    \hat U = \mathcal{T}e^{-\ii\int\dd t\,\hat H^t_I(t)}\,.
\end{align}
In general we can evaluate this perturbatively via Dyson series expansion
\begin{align}
    \hat U &= \openone + \hat U^{(1)}+ \hat U^{(2)}+O(\lambda^3)\,,\\
    \hat U^{(1)} &= -\ii\int_{-\infty}^\infty \dd t\,\hat H_I^t(t)\,,\\
    \hat U^{(2)} &= -\int_{-\infty}^\infty \dd t\int_{-\infty}^t\dd t' \, \hat H_I^t(t)\hat H_I^{t}(t')\,.
\end{align}
Thus the final state of the full system can be described by perturbative Dyson expansion about the initial state:
\begin{align}
    \hat\rho &= \hat{\rho}_0 + \hat{\rho}^{(1)} + \hat{\rho}^{(2)} + O(\lambda^3)\,,\\
    \hat{\rho}^{(j)} &= \sum_{k+l=j} \hat U^{(k)}\hat{\rho}_0 \hat U^{(l)\dagger}\,,
    \label{eq: Dyson-series}
\end{align}
where $\hat{\rho}^{(j)}$ is of order $\lambda^j$. The final state of the two detectors are obtained by tracing out the field, thus we also have a perturbative expansion
\begin{align}
    \hat{\rho}_{\ab} &= \tr_{\phi} \hat\rho = \hat{\rho}_{\ab,0} +\hat{\rho}_{\ab}^{(1)} + \hat{\rho}_{\ab}^{(2)}+O(\lambda^3)\,,
\end{align}
where $\hat{\rho}_{\ab}^{(j)} =\tr_\phi\hat{\rho}^{(j)}$ and $\hat{\rho}_{\ab,0} =\tr_\phi\hat{\rho}_0$.  

If the initial state of the field is the vacuum state $\ket{0}$ defined by $\hat a_\bk\ket 0 = 0$ for all $\bk$, then $\hat{\rho}_\ab^{(1)}=0$ due to the vanishing of the one-point function $\braket{0|\hat\phi(\sx)|0}$. Thus the leading order correction to the joint bipartite density matrix $\hat{\rho}_{\ab,0}$ is of order $\lambda^2$. 

For the purpose of analysing entanglement harvesting protocol, we will also make the assumption that both detectors are initially uncorrelated and are in their own respective ground states {with respect to their free Hamiltonian}, thus we write
\begin{align}
    \hat{\rho}_{0} = \ket{g_\textsc{a}}\!\bra{g_\textsc{a}}\otimes\ket{g_\textsc{b}}\!\bra{g_\textsc{b}}\otimes \ket{0}\!\bra{0}\,.
\end{align}
For simplicity we consider both detectors to be static relative to the quantization frame so that the coordinates of their trajectories are given by $(t(\tau_j),\bx(\tau_j)) = (\tau_j,\bx_j)$ for some fixed $\bx_j$. Under these assumptions, we can show that to leading order and in the ordered basis $\{\ket{g_\textsc{a} g_\textsc{b}},\ket{g_\textsc{a} e_\textsc{b}},\ket{e_\textsc{a} g_\textsc{b}},\ket{e_\textsc{a} e_\textsc{b}}\}$ we get
\begin{align}
    \hat{\rho}_{\ab} &= 
    \begin{pmatrix}
        1-\mathcal{L}_\textsc{aa}-\mathcal{L}_\textsc{bb} & 0 & 0 & \mathcal{M}^* \\
        0 & \mathcal{L}_\textsc{bb} & \mathcal{L}_\textsc{ab} & 0 \\
        0 & \mathcal{L}_\textsc{ba} & \mathcal{L}_\textsc{aa} & 0 \\
        \mathcal{M} & 0 & 0 & 0 
    \end{pmatrix}+O(\lambda^4)\,,
    \label{eq: final-detector-matrix}
\end{align}
where the matrix elements are given by
\begin{align}
    \mathcal{L}_{ij} &= \lambda^2\int \dd t\,\dd t'\,\chi_i(t)\chi_j(t') e^{-\ii\Omega(t - t')}W(t,{\bx_i} ;t',\bx_j)
    \label{eq: Lij}\\
    \mathcal{M} &= -\lambda^2\int \dd t\,\dd t'\, e^{\ii\Omega(t+t')} \chi_\textsc{a}(t)\chi_\textsc{b}(t')\notag\\
    &\phantom{=}\hspace{1.75cm}\times \bigr[ \Theta(t-t') W(t,\bx_\textsc{a};t',\bx_\textsc{b})\notag\\
    &\phantom{=}\hspace{1.75cm}+\Theta(t' - t)W(t',\bx_\textsc{b};t,\bx_\textsc{a})\bigr]\,,
    \label{eq: M-nonloc}
\end{align}
$W(\sx_i(\tau_i),\sx_j(\tau_j'))$ is the pullback of the Wightman function along the detectors' trajectories and $\Theta(z)$ is the Heaviside function.

In order to measure the amount of entanglement between the two qubits, we use  entanglement measures such as negativity or concurrence \cite{Vidal2002negativity,Wotters1998entanglementmeasure,Horodecki996separable}. For a system of two qubits, the negativity $\mathcal{N}$ for the density matrix $\hat\rho$ is a faithful entanglement monotone defined by \cite{Vidal2002negativity}
\begin{align}
    \mathcal{N}[\hat\rho]\coloneqq \frac{\left|\left|{\hat\rho}^\Gamma\right|\right|_1-1}{2}\,,
\end{align}
where $\hat\rho^\Gamma$ is the partial transpose of $\hat\rho$ and $\left|\left|\cdot\right|\right|_1$ is the trace norm. For the final density matrix $\hat{\rho}_\ab$ in Eq.~\eqref{eq: final-detector-matrix}, negativity takes the form
\begin{align}
    \mathcal{N}[\hat{\rho}_\ab]&= \max\{0, -E\} + O(\lambda^4)\,,
\end{align}
where
\begin{align}
    E&= \frac{1}{2}\rr{\mathcal{L}_\textsc{aa}+\mathcal{L}_\textsc{bb}-\sqrt{(\mathcal{L}_\textsc{aa}-\mathcal{L}_\textsc{bb})^2+4|\mathcal{M}|^2}}\,.
\end{align}
Since the detectors are identical and Minkowski space has translational symmetries, we have that $\mathcal{L}_\textsc{aa}=\mathcal{L}_\textsc{bb}$ and hence the negativity reduces to
\begin{align}
    \mathcal{N}[\hat{\rho}_\ab]\coloneqq \max\left\{0,|\mathcal{M}|-\mathcal{L}_{jj}\right\} + O(\lambda^4)\,.
\end{align}

\section{Wightman function and strong Huygens' principle}
\label{sec: setup}

For our purposes, we need to analyze the (vacuum) two-point Wightman function $W(\sx,\sx')=\braket{0|\hat\phi(\sx)\hat\phi(\sx')|0}$. For $(n+1)$-dimensional Minkowski spacetime, this reads
\begin{align}
    W(\sx,\sx')
    &= \int\frac{\dd^n\bk}{2(2\pi)^n\omega_\bk}e^{-\ii\omega_\bk(t-t')+\ii\bk\cdot(\bx-\bx')}\,,
\end{align}
where it is understood that the Wightman function is a (bi-)distribution.

The Wightman function has two clearly distinct contributions, its imaginary and real parts:
\begin{align}
    W(\sx,\sx') &\coloneqq \frac{1}{2}\rr{C^+(\sx,\sx')+ C^-(\sx,\sx')}\,,\label{eq: commutator-anticomm}
\end{align}
where 
\begin{align}
    C^+(\sx,\sx') &= \braket{0|\{\hat\phi(\sx),\hat\phi(\sx')\}|0} \,,
    \label{eq: anti-comm}\\
    C^-(\sx,\sx') &= \braket{0|[\hat\phi(\sx),\hat\phi(\sx')]|0} \,.
    \label{eq: comm}
\end{align}
This splitting is motivated by two important facts on which the main crux of this work is based on: 
\begin{enumerate}[label=(\roman*),leftmargin=*]
    \item The expectation value of the field commutator $[\hat\phi(\sx),\hat\phi(\sx')]$ is \textit{state-independent}: that is, if $\hat\rho_\phi,\hat\rho_\phi'$ are two distinct field states and $C^-,{C'}^-$ their corresponding commutator expectation values, then
    \begin{align}
        C^-(\sx,\sx') &= \tr\rr{\hat\rho_\phi[\hat\phi(\sx),\hat\phi(\sx')]} \notag\\
        &= \tr\rr{\hat\rho'_\phi[\hat\phi(\sx),\hat\phi(\sx')]} \notag\\
        &= {C'}^-(\sx,\sx')\,.
    \end{align}
    In particular, it means that $C^-(\sx,\sx')$ for vacuum state will be the same as the one computed using a field state which has \textit{no correlations whatsoever}.
    
    \item The expectation value of the anti-commutator $\{\hat\phi(\sx),\hat\phi(\sx')\}$ is \textit{state-dependent}: that is, if $\hat\rho_\phi,\hat\rho_{\phi}'$  are two distinct field states and $C^+,{C'}^+$ their corresponding anti-commutator expectation values, then in general
    \begin{align}
        C^+(\sx,\sx') &=  \tr\rr{\hat\rho_\phi\{\hat\phi(\sx),\hat\phi(\sx')\}} \notag\\
        &\neq \tr\rr{\hat\rho'_\phi\{\hat\phi(\sx),\hat\phi(\sx')\}} \notag\\
        &= {C'}^+(\sx,\sx')\,.
    \end{align}
    In particular, it means that the difference in the two-point correlations between two field states is completely contained in the expectation value of the anti-commutator.
\end{enumerate}
We also note that field commutator \eqref{eq: comm} is a Green's function for the Klein-Gordon equation in $(n+1)$-dimensional Minkowski spacetime. 



The decomposition into commutator and anti-commutator is very helpful to disentangle entanglement harvesting (no pun intended) from the entanglement that is not harvested, but rather generated through field-mediated communication of the two detectors. Since the field commutator is state-independent, the bipartite entanglement of the detectors cannot be associated to pre-existing (vacuum) correlations of the field. It is also known that communication between the detectors is, at leading order, given by the field commutator~\cite{Jonsson2014cavityQED,Causality2015Eduardo,Blasco2015Huygens,Casals2020commBH,Simidzija2020capacity}.

In order to better understand the role of communication in generating entanglement between two detectors, we need some results about classical Green's functions for wave propagation. The \textit{strong Huygens' principle} states that the Green's functions (hence the general solutions) of a second-order linear partial differential equation of normal hyperbolic type has support only along the null direction (the boundary of the domain of dependence, e.g., the light cone) \cite{McLenaghan1974huygens}. For a massless Klein-Gordon field in flat spacetimes, a classic result shows that this wave equation satisfies the strong Huygens' principle for odd $n\geq 3$~\cite{CourantHilbert1989Vol2ch6}. When the principle is violated, the Green's function also has support in the interior of the light cone. The principle is known not to hold for massless fields in generic curved spacetimes and for fields with nonzero mass~\cite{Sonego1992huygenscurved,Faraoni2019huygens}. 

The preceding discussion shows that the support of the field commutator $C^-(\sx,\sx')$ is paramount in determining the role of communication between two detectors when they are causally connected since it mediates the leading order communication. In contrast, the anti-commutator $C^+(\sx,\sx')$ has support for spacelike-separated events and it is certainly the only contribution to the Wightman function (and therefore the correlations that two detectors can acquire) when two detectors perform spacelike entanglement harvesting.  

In the next section we will build an estimator of how much of the entanglement acquired between two detectors is due to communication (which is to say, through the field commutator) and how much is coming through the anti-commutator, which plays no role in communication at leading order~\cite{Jonsson2014cavityQED,Casals2020commBH,Blasco2015Huygens}. The contribution from the anti-commutator will therefore be associated with genuine harvesting of entanglement. 

To carry out this study this we will generalize the entanglement harvesting protocol in~\cite{pozas2015harvesting} to arbitrary $(n+1)$-dimensional spacetimes and also for scalar fields with $m>0$.

\section{Communication and entanglement harvesting in arbitrary dimensions}
\label{sec: comm}

Our first task is to obtain explicit expressions for the matrix elements $\mathcal{L}_{\textsc{aa}}$ ($\mathcal{L}_{\textsc{bb}}$) and $\mathcal{M}$ for arbitrary field mass $m$ and any number of spacetime dimensions. 

\begin{figure}[tp]
    \centering
    \includegraphics[scale=0.67]{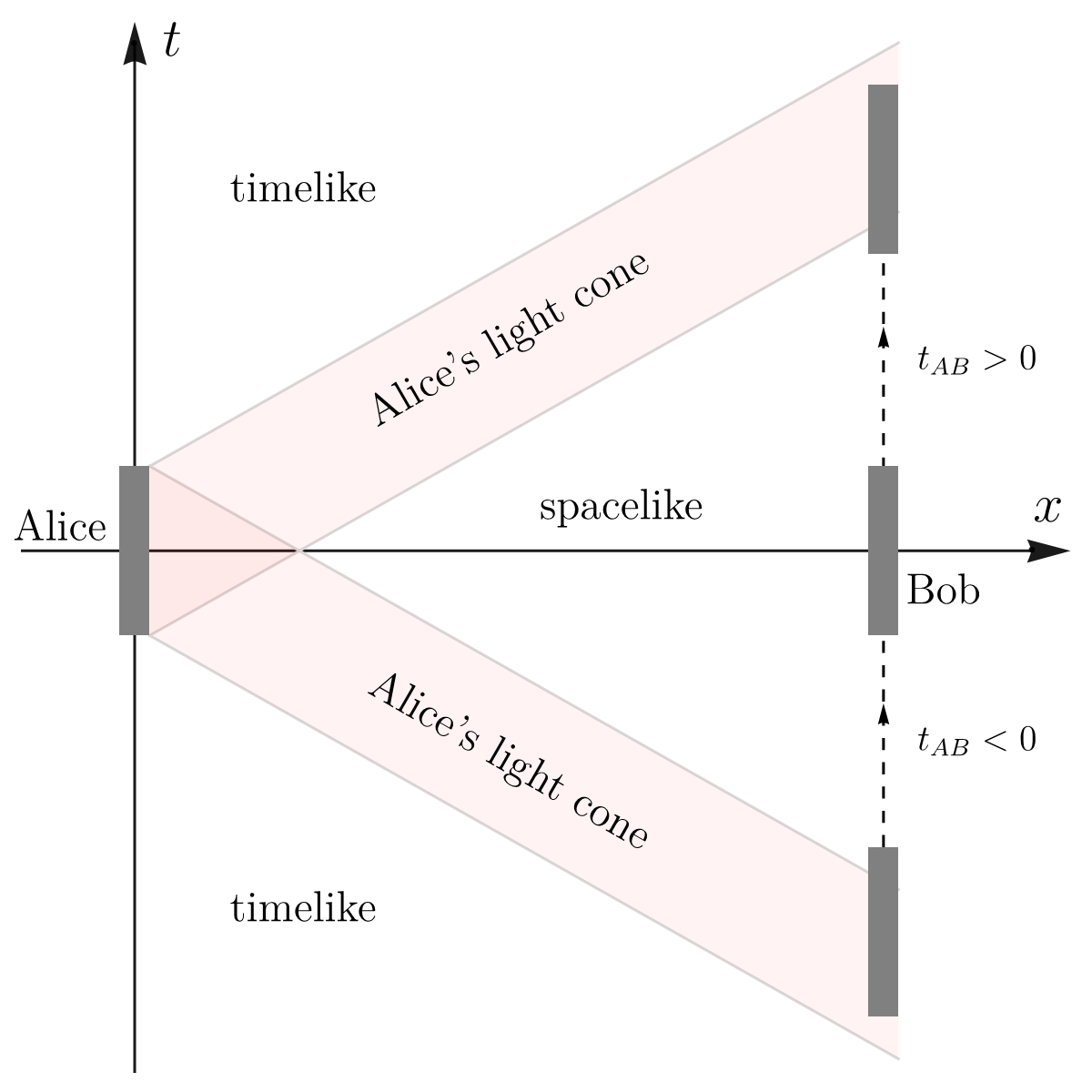}
    \caption{\textbf{Spacetime diagram for Alice and Bob's detectors.} The grey rectangles are the (strong) support of their detectors' switching functions, denoted $S_\textsc{a},S_\textsc{b}$. Alice and Bob are separated by proper distance $L$. The time delay $t_{\ab}=t_\textsc{b}-t_\textsc{a}$ marks the difference between their switching peaks. The red shaded regions are null-separated from $S_\textsc{a}$.}
    \label{fig: strong-supp-causal}
\end{figure}

Let us take the switching function $j$ to be Gaussian 
\begin{align}
    \chi_j(t) &= e^{-\frac{(t-t_j)^2}{T^2}}\,,\quad j=\textsc{A,B}\,, 
    \label{eq: Gaussian-switch}
\end{align}
where $T$ prescribes the effective duration of the interaction and $t_j$ denotes the switching peak of detector $j$. With this choice, the matrix elements of $\hat{\rho}_\ab$ will greatly simplify. In this work, we define the \textit{strong support} of the detectors to be the interval 
\begin{align}
    S_j = [-3.5T+t_j,3.5T+t_j]\,, \quad j = \textsc{A,B}\,,
    \label{eq: strong support}
\end{align}
which contains $99.9999\%$ of the total area of the Gaussian\footnote{As we will see in Section~\ref{sec: results}, the choice of $\pm 3.5T$ about the centre of Gaussian is based on numerical evidence involving the field commutator. It also suggests that the interval for the strong support $S_j$ should not be taken to be smaller than \eqref{eq: strong support}. In any case, we will study truly compactly supported switching functions in more detail in Section~\ref{sec: further-results}.}. This allows us to think of the switching as effectively compactly supported within an interval of $7T$ centered at $t_j$. Detector B can then be considered  spacelike separated from detector A when $S_\textsc{b}$ does not intersect any light rays emanating from $S_\textsc{a}$, as we show schematically in Figure~\ref{fig: strong-supp-causal}.

The matrix element $\mathcal{L}_{jj}$, which corresponds to the vacuum excitation probability of detector $j$, is given by
\begin{align}
    \mathcal{L}_{jj} &= \lambda^2\int \frac{\dd^n\bk}{2(2\pi)^n\omega_{\bk}}|\tilde\chi_j(\Omega+{\omega_\bk})|^2\,,
    \label{eq: probability}
\end{align}
where $\tilde\chi$ is the Fourier transform of the switching function.  For a massless scalar field with $\omega_\bk=|\bk|$ and Gaussian switching \eqref{eq: Gaussian-switch}, this can be solved exactly:
\begin{align}
    \mathcal{L}_{jj} &= \frac{ \pi ^{\frac{2-n}{2}} T^{3-n}}{2^{\frac{n+3}{2}}\Gamma \left(\frac{n}{2}\right)}\bigg[\Gamma \left(\frac{n-1}{2}\right) {_1F_1}\left(\frac{2-n}{2};\frac{1}{2};-\frac{ T^2 \Omega ^2 }{2}\right)\notag\\
    &\hspace{0.25cm}-
    \sqrt{2} T \Omega \, \Gamma\left(\frac{n}{2}\right) {_1F_1}\left(\frac{3-n}{2};\frac{3}{2};-\frac{T^2 \Omega ^2}{2} \right)\bigg]\,,
\end{align}
where ${_1F_1}(a;b;z)$ is Kummer's confluent hypergeometric function and $\Gamma(z)$ is the gamma function \cite{NIST:DLMF,abramowitz1965handbook}. This expression is valid for $n>1$ since there is a well-known infrared (IR) divergence in (1+1) dimensions\footnote{If we were to continue using this expression for $(1+1)D$ case, then one should use $n=1+\epsilon$ for some $0<\epsilon\ll1$, which amounts to dimensional regularization of the IR divergence. One can also use mass regularization (small non-zero mass) or a hard IR cutoff as in~\cite{pozas2015harvesting}.} \cite{birrell1984QFTCS,pozas2015harvesting}. For massive scalar fields where $\omega_\bk=\sqrt{|\bk^2|+m^2}$ with $m>0$, there is no closed form expression for \eqref{eq: probability}. 

For matrix element $\mathcal{M}$ which depends on the trajectories of both detectors, we decompose it into two parts
\begin{align}
    \mathcal{M} = \mathcal{M}^++\mathcal{M}^-\,,
    \label{eq: M-splitting}
\end{align}
where $\mathcal{M}^\pm$ depends on the (anti-)commutator $C^\pm(\sx,\sx')$ in Eqs.~\eqref{eq: anti-comm} and \eqref{eq: comm}. Using the shorthand $k\equiv |\bk|$, they are given by (see Appendix~\ref{appendix: M-proof})
\begin{align}
    \mathcal{M}^+ &= -\lambda^2e^{2\ii\Omega t_\textsc{a}}\int_0^\infty\!\!\! {\frac{\dd k\,k^{n-1}}{\sqrt{k^2+m^2}}}\rr{\mathcal{K}_1(k)+\mathcal{K}_2(k)}\,,
    \label{eq: Mplus}\\
    \mathcal{M}^- &= -\lambda^2e^{2\ii\Omega t_\textsc{a}}\int_0^\infty\!\!\! {\frac{\dd k\,k^{n-1}}{\sqrt{k^2+m^2}}}\rr{\mathcal{K}_3(k)+\mathcal{K}_4(k)}
    \,.
    \label{eq: Mmin}
\end{align}
where each $\mathcal{K}_j$ ($j=1,2,3,4$) reads
\begin{align}
    \mathcal{K}_1(k) &= 2^{-n-1} \pi ^{1-\frac{n}{2}} T^2 \, _0\tilde{F}_1\left(\frac{n}{2};-\frac{k^2L^2}{4} \right)\notag\\
    &\times e^{-\frac{1}{2} T^2 \left(k^2+\Omega ^2\right)+ \ii t_{\ab}  (\Omega -k)}\,,\\
    \mathcal{K}_2(k) &= 2^{-n-1} \pi ^{1-\frac{n}{2}} T^2 \, _0\tilde{F}_1\left(\frac{n}{2};-\frac{k^2L^2}{4}\right)\notag\\
    &\times e^{-\frac{1}{2} T^2 \left(k^2+\Omega ^2\right)+ \ii t_{\ab}  (k+\Omega )}\,,\\
    \mathcal{K}_3(k) &= -\ii 2^{-n} \pi ^{\frac{1-n}{2}} T^2 e^{\ii t_{\ab}  \Omega -\frac{t_{\ab} ^2}{2 T^2}-\frac{T^2 \Omega ^2}{2}} \mathcal{F}\left(\frac{k T^2+\ii t_{\ab} }{\sqrt{2} T}\right)\notag\\
    &\times \, _0\tilde{F}_1\left(\frac{n}{2};-\frac{k^2L^2}{4}\right) \,,\\
    \mathcal{K}_4(k) &= -\ii 2^{-n} \pi ^{\frac{1-n}{2}} T^2 e^{\ii t_{\ab}  \Omega -\frac{t_{\ab} ^2}{2 T^2}-\frac{T^2 \Omega ^2}{2}} \mathcal{F}\left(\frac{k T^2-\ii t_{\ab} }{\sqrt{2} T}\right) \notag\\
    &\times\, _0\tilde{F}_1\left(\frac{n}{2};-\frac{k^2L^2}{4} \right)\,,
\end{align}
where $\mathcal{F}(z)=e^{-z^2}\int_0^z\dd y\,e^{z^2}$ is the Dawson's integral and  ${_p\tilde{F}_q(b;z)}$ is the regularized generalized hypergeometric function or Bessel-Clifford function\footnote{The non-regularized, generalized hypergeometric function is related to the regularized one by $\Gamma(b){_p\tilde{F}_q(b;z)} = {_p{F}_q(a;z)}$ \cite{Weisstein2021reghyp}. Note that another commonly used expression for ${_0\tilde{F}_1}$ involves the Bessel function of the first kind, often called the Bessel-Clifford function $\mathcal{C}_n$. They are related by $\mathcal{C}_n(-z^2/4)\equiv {_0\tilde{F}_1(n+1;-z^2/4)} = (2/z)^{n}J_{n}(z)$ \cite{abramowitz1965handbook}.} \cite{Weisstein2021reghyp,NIST:DLMF,abramowitz1965handbook}. Here we use the shorthand $t_{\ab}\coloneqq t_\textsc{b}-t_\textsc{a}$ for the time delay. As there is no closed form expressions for $\mathcal{M}$ for arbitrary $m$ and $t_{\ab}$, we will evaluate $\mathcal{M}$ numerically.

The splitting in Eq.~\eqref{eq: M-splitting} motivates us to define \textit{harvested negativity} $\mathcal{N}^+[\hat{\rho}_{\ab}]$ and \textit{communication-assisted negativity} $\mathcal{N}^-[\hat{\rho}_{\ab}] $ as
\begin{align}
    \mathcal{N}^\pm[\hat{\rho}_{\ab}] &\coloneqq  \max\left\{0,|\mathcal{M}^\pm|-\mathcal{L}_{jj}\right\} + O(\lambda^4)\,.
\end{align}
The idea is that if the two detectors are spacelike separated, then $\mathcal{M}^- =  0$ and hence $\mathcal{N} = \mathcal{N}^+$ ($\mathcal{N}^- \!= 0$). When the detectors are not in spacelike separation, they still can in principle harvest entanglement. Indeed the vacuum in any two regions of spacetime contain quantum correlations~\cite{summers1985bell,summers1987bell,Higuchi2017wedges}. Comparing the contributions of the commutator and anti-commutator to negativity will hence allow us to see how much of the  entanglement between the detectors is due to bipartite communication and how much is actually harvested from the scalar field vacuum.

\begin{figure*}[tp]
    \centering
    \includegraphics[scale=0.64]{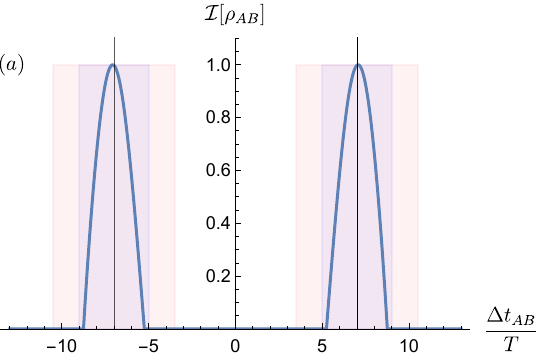}
    \includegraphics[scale=0.64]{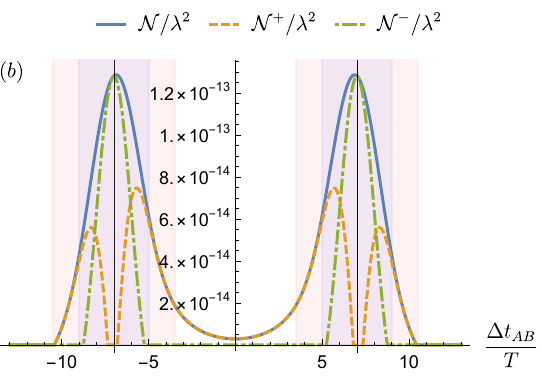}
    \includegraphics[scale=0.64]{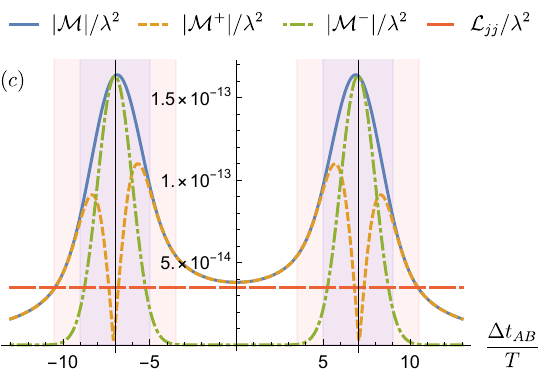}
    \caption{\textbf{Bipartite entanglement as a function of time delay} $t_\ab$ \textbf{between Alice and Bob's switching in (3+1) dimensions.} The parameters are $\Omega T = 7$ and $L=7T$. The vertical straight lines are the light cones of detector $A$ emanating from the event $(t_\textsc{a},\bm{0})$. The red shaded region marks the strong support of Alice's switching function, and the blue-shaded area marks the region where the behaviour of $|\mathcal{M}^\pm|$ starts to change dramatically. (a) The communication-assisted entanglement estimator. Note that $\mathcal{I}[\hat{\rho}_\ab]\approx 1$ near the light cone, hence \textit{most} of the bipartite entanglement is purely communication-based. (b) $\mathcal{N},\mathcal{N}^\pm$ as a function of $t_\ab$. Crucially, the anti-commutator part $|\mathcal{M}^+|$ vanishes near the light cone while the commutator part $|\mathcal{M}^-|$ dominates.  (c) $|\mathcal{M}|,|\mathcal{M}^\pm|,\mathcal{L}_{jj}$ as a function of $t_\ab$. The region where $|\mathcal{M}|> \mathcal{L}_{jj}$ (solid blue curve is above dashed horizontal red curve) is where the negativity $\mathcal{N}$ is nonzero. }
    \label{fig: concurrence3D-1}
\end{figure*}

To compare both contributions, we define a \textit{communication-mediated entanglement estimator} $\mathcal I[\hat{\rho}_\ab]$ given by
\begin{align}
    \mathcal{I}[\hat{\rho}_{\ab}]\coloneqq
    \begin{cases}
    \displaystyle
    \frac{\mathcal N^-[\hat{\rho}_{\ab}]}{\mathcal N[\hat{\rho}_{\ab}]} \hspace{0.5cm} &\mathcal N[\hat{\rho}_{\ab}]>0\\
    0   &\mathcal N[\hat{\rho}_{\ab}]=0
    \end{cases}
    \label{eq: communication-estimator}
\end{align}
The estimator's role can be summarized as follows: 
\begin{itemize}[leftmargin=*]
    \item If $\mathcal{I}[\hat{\rho}_\ab] \approx 1$, then essentially all of the entanglement is dominated by the communication between two detectors through the field and not from swapping entanglement with the scalar field vacuum.
    
    \item If $0 < \mathcal{I}[\hat{\rho}_\ab]< 1$, then some of the entanglement is communication-assisted, and vacuum entanglement also has nonzero contribution to the detector-detector entanglement.
    
    \item If $\mathcal{I}[\hat{\rho}_\ab] = 0$ then either there is no entanglement ($\mathcal N[\hat{\rho}_\ab]=0$) or all entanglement comes from harvesting $(\mathcal N^-[\hat{\rho}_\ab] = 0)$ since the anti-commutator does not participate at all in leading order communication~\cite{Jonsson2014cavityQED,Causality2015Eduardo,Casals2020commBH}. These two cases can be distinguished by checking whether $|\mathcal{M}|>\mathcal{L}_{jj}$.
    
\end{itemize}
We will show in the next section that the estimator can attain values close to unity when the detectors are in causal contact.

In what follows, we are going to focus on varying only the time delay between the switching peaks $t_\ab$. In particular, the variation of $t_\ab$ will allow us to change the causal relationships between detector $A$ and $B$. All quantities will be measured in units of the Gaussian switching width $T$. For concreteness, we will set the proper distance between Alice and Bob's detectors to be some fixed quantity $\Omega T = 7$ and $L= 7T$. In making these choices, one important thing is that $L$ be sufficiently large so that the strong support \eqref{eq: strong support} still gives enough space between detectors for spacelike separation to be well-defined. 

Moreover, the calculations done in this work can be straightforwardly extended to the case when the detectors have finite size: the inclusion of spatial smearing is outlined in Appendix~\ref{appendix: spatial smearing}. We focus on pointlike detectors so that the causal relationships between the two detectors are clearer as it is completely controlled by the switching function.

Finally, we re-emphasize that even though we are working with Gaussian switching in this section and the next one, and hence the detectors are really never truly spacelike separated, we will show in Section~\ref{sec: further-results} that the results carry to the case of strictly compactly supported switching. In other words, the negligible Gaussian tails outside of the detectors' switching strong support have no relevance to entanglement harvesting in general, and in particular to our results.

\section{Results}
\label{sec: results}
In this section we show the result for $(3+1),(2+1)$ and $(1+1)$ dimensions when the scalar field is massless and the switching is Gaussian. We will consider higher dimensions, massive fields and compactly supported switching functions in Section~\ref{sec: further-results}.

\subsection{(3+1) dimensions} 
In Figure~\ref{fig: concurrence3D-1}, we plot the communication-assisted entanglement estimator $\skri[\hat{\rho}_{\ab}]$, the negativity and the matrix elements $\mathcal{M},\mathcal{L}_{jj}$ for (3+1) dimensions.
The vertical straight lines are the light cones of detector A emanating from the event $(t_\textsc{a},\bm{0})$, and we vary the time delay $t_\ab$. In Figure~\ref{fig: concurrence3D-1}(b) we show the total negativity $\mathcal{N}$ of the two detectors after interaction as well as the decomposition into harvested and communication-assisted negativity $\mathcal{N}^\pm$. In Figure~\ref{fig: concurrence3D-1}(c) we show in more detail the behaviour of the matrix elements of $\hat{\rho}_{\ab}$. For all figures, the red-colored shaded area marks Alice's light cone with respect to the strong support $S_\textsc{a}$ (cf. Figure~\ref{fig: strong-supp-causal}). The blue-shaded area marks the region where the behaviour of $|\mathcal{M}^\pm|$ starts to change dramatically, which occurs within Alice's light cone. The central white area about the origin is where Alice and Bob are effectively spacelike separated, as one can verify by checking the commutator-dependent quantities $\mathcal{N}^-$ and $|\mathcal{M}^-|$ in Figure~\ref{fig: concurrence3D-1}(b,c).

From Figure~\ref{fig: concurrence3D-1}(a), we see that in (3+1) dimensions, the communication-assisted entanglement estimator \mbox{$\mathcal{I}[\hat{\rho}_\ab]\approx 1$} near the light cone at $t_\ab=\pm 7T$ (since $L=7T$). This means that essentially all of the bipartite entanglement is communication-based and not harvested from the scalar field vacuum. Figure~\ref{fig: concurrence3D-1}(b) and~\ref{fig: concurrence3D-1}(c) show how the anti-commutator (state-dependent) part takes a sudden, drastic dip (near the edges of the blue shaded region) as full light-contact is approached, eventually vanishing at the light cone; in contrast, the commutator part starts to dominate at precisely the regions where the anti-commutator contribution starts diminishing.

From the field-theoretic perspective, this result may perhaps be somewhat surprising because it says that communication does \textit{not} simply enhance bipartite entanglement between Alice and Bob by ``adding'' more correlations on top of vacuum entanglement harvesting. Even though a Bogoliubov decomposition analysis shows that timelike separated regions do contain correlations~\cite{Higuchi2017wedges}, our results suggest that when the detectors can communicate through the field, the two detectors will \textit{forgo} entanglement harvesting from the vacuum and preferentially gain entanglement through their exchange of information through the field. Indeed, we emphasize that since the commutator contribution is \textit{state-independent}, any entanglement obtained by the detectors from the commutator cannot be attributed to pre-existing correlations of the vacuum state of the field.

The fact that the peaks in $\mathcal{I}[\hat{\rho}_\ab]$ are localized around the light cone is a consequence of the strong Huygen's principle in (3+1) dimensions: the (expectation value of) commutator  $[\phi(\sx),\phi(\sx')]$ for massless field only has support along the null direction. The explicit expression reads (see, e.g., Appendix~\ref{appendix: strong-Huygens} for a derivation)
\begin{align}
    C^-_3(\sx,\sx') &= \frac{\ii}{4\pi |\Delta \bx|}\left[\delta(\Delta t + |\Delta\bx|) - \delta(\Delta t - |\Delta\bx|)\right]\,,
    \label{eq: commutator-3D}
\end{align}
where $\delta(z)$ is a one-dimensional Dirac delta distribution and we used the notation $C^-_{n}$ to denote the commutator in arbitrary $(n+1)$-dimensional Minkowski spacetime.

Next, we note that when the detectors are timelike separated, it is in principle possible to have \textit{timelike entanglement harvesting} as the field commutator completely vanishes outside the light cone, while the anti-commutator still has support in the light cone interior (see e.g. \cite{olson2012timelikeharvesting} for related result). However, it is generically much more difficult to extract entanglement from the vacuum for timelike separation than for spacelike separation (for fixed proper separation $L$). This follows naturally from the fact that the Wightman function for massless fields in (3+1) dimensions has a power law decay $\sigma(\sx_\textsc{a},\sx_\textsc{b})^{-1}$, where $\sigma(\sx,\sy)$ is the Synge world function, which in flat space reduces to half the spacetime interval:
\begin{align}
    \sigma(\sx,\sy) = \frac{1}{2}\rr{|x^0-y^0|^2-|\bx-\by|^2}\,.
\end{align}
Since the commutator is supported only at the light cone, it follows that this power law falloff is contained in the anti-commutator. Therefore, the anti-commutator contribution $|\mathcal{M}^+|$ diminishes the deeper Bob is in Alice's light cone interior, eventually falling below the noise term $\mathcal{L}_{jj}$ rendering harvesting impossible. 

Let us comment on one minor observation concerning the slight asymmetry of the estimator $\mathcal{I}[\hat{\rho}_\ab]$ in Figure~\ref{fig: concurrence3D-1}. The peaks of $\mathcal{I}[\hat{\rho}_\ab]$ is not exactly at $\Delta t_\ab=7T$ (the light cone emanating from the peak of Alice's Gaussian switching) but comes very close to it. This has to do with the inherent asymmetry of the anti-commutator contribution $|\mathcal{M}^+|$ (see Figure~\ref{fig: concurrence3D-1}(c)) that affects the denominator of the ratio of $\mathcal{N}^-/\mathcal{N}$ in Eq.~\eqref{eq: communication-estimator}. One can check numerically that for the parameters we chose in Figure~\ref{fig: concurrence3D-1}, the value of $|\mathcal{M}^+|$ vanishes at approximately $\Delta t_\ab\approx \pm 7.07 T$. In contrast, the commutator contribution $|\mathcal{M}^-|$ is indeed symmetric about the light cone. Note that the symmetry of $|\mathcal{M}^-|$ only occurs at (3+1) dimensions and has no direct connection with null separation: we will see in Section~\ref{sec: further-results} that in higher dimensions the asymmetry manifests also for $|\mathcal{M}^-|$ regardless of the strong Huygens' principle. In any case the finite nature of the switching function blurs the picture, and What really matters is that in the neighbourhood of $\Delta t_\ab=7T$ (blue region of Figure~\ref{fig: concurrence3D-1}), the bipartite entanglement is dominated by field-mediated communication.

In summary, our result in (3+1) dimensions highlights the importance of the detectors being spacelike separated in order for vacuum entanglement harvesting to be possible. When they are null separated, the entanglement comes mainly from bipartite communication and not from entanglement harvesting. When they are timelike separated, entanglement harvesting is in principle possible but much more difficult  than spacelike harvesting due to the power-law decay of the anti-commutator.

\begin{figure*}[htp]
    \centering
    \includegraphics[scale=0.64]{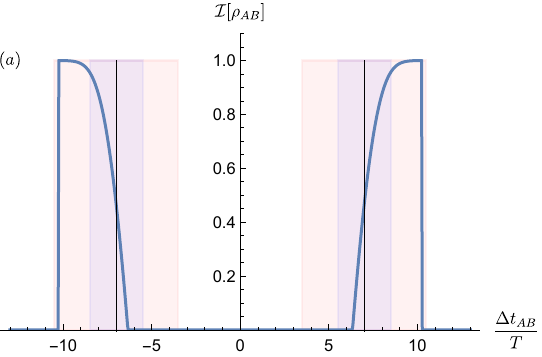}
    \includegraphics[scale=0.64]{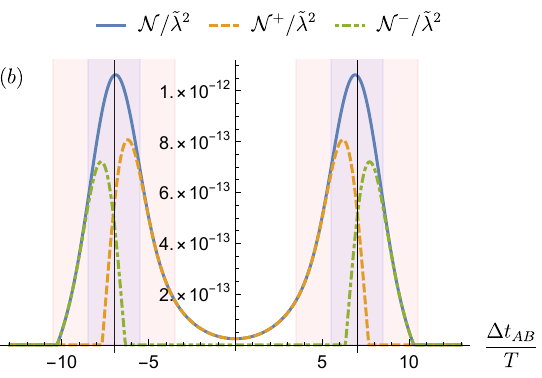}
    \includegraphics[scale=0.64]{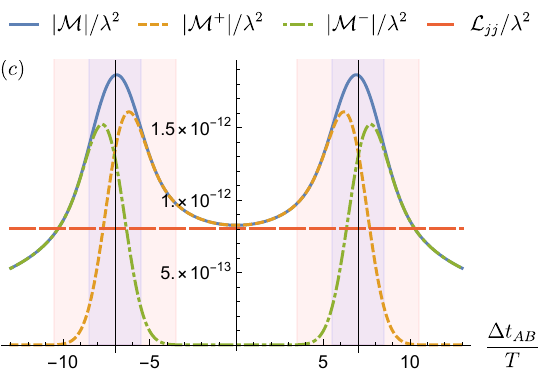}
    \caption{\textbf{Bipartite entanglement as a function of time delay} $t_\textsc{ab}$ \textbf{between Alice and Bob's switching in (2+1) dimensions.} The parameters are $\Omega T = 7$ and $L=7T$. The vertical straight lines are the light cones of detector $A$ emanating from the event $(t_\textsc{a},\bm{0})$. The shaded region marks the strong support of Alice's switching function, and the blue-shaded area marks the region where the behaviour of $|\mathcal{M}^\pm|$ starts to change dramatically.  (a) The communication-assisted entanglement estimator. Note that $\mathcal{I}[\hat{\rho}_\ab]\approx 1$ near the light cone, hence \textit{all} of the bipartite entanglement is purely communication-based. (b) $\mathcal{N},\mathcal{N}^\pm$ as a function of $t_\ab$. Crucially, the anti-commutator part $|\mathcal{M}^+|$ vanishes near the light cone while the commutator part $|\mathcal{M}^-|$ dominates.  (c) $|\mathcal{M}|,|\mathcal{M}^\pm|,\mathcal{L}_{jj}$ as a function of $t_\ab$. The region where $|\mathcal{M}|> \mathcal{L}_{jj}$ (solid blue curve is above dashed horizontal red curve) is where the negativity $\mathcal{N}$ is nonzero.}
    \label{fig: concurrence2D-1}
\end{figure*}

\subsection{(2+1) dimensions}

Let us now see what happens in (2+1) dimensions where the strong Huygens' principle is known to not hold, as we show in Figure~\ref{fig: concurrence2D-1}. Note that since $\lambda$ has units of $[\text{Length}]^{\frac{n-3}{2}}$ in natural units, we define the dimensionless coupling constant $\tilde\lambda=\lambda T^{\frac{3-n}{2}}$ since $T$ is fixed in this work.

We see in Figure~\ref{fig: concurrence2D-1}(a) that as Bob enters deeper into the interior of Alice's light cone, the communication-assisted entanglement estimator  $\mathcal{I}[\hat{\rho}_\ab]\to 1$. The sudden vanishing of $\mathcal{I}[\hat{\rho}_\ab]$ for $|t_\ab|\gtrsim 10T$ is just because there is no more entanglement past this point: $\mathcal{M}^\pm\to 0$ as $|t_\ab|\to \infty$ (while $\mathcal{L}_{jj}$ remains constant), which follows from the falloff properties of the Wightman function for $n\geq 2$. Inspection of Figures~\ref{fig: concurrence2D-1}(b) and (c) shows that within Alice's light cone interior, we have that $|\mathcal{M}|\approx |\mathcal{M}^-|$, thus any entanglement generated in the timelike region is all communication-based: there is virtually no entanglement harvesting for timelike separated detectors. On the other hand, unlike the (3+1) dimensional case, the negativity at null-separation is shared equally by communication and harvesting at the light cone. Furthermore, the violation of the strong Huygens' principle manifests itself by having the field commutator slowly increasing its dominance as Bob approaches Alice's light cone, eventually taking over all of $|\mathcal{M}|\approx | \mathcal{M}^-|$. At the same time the role of the anti-commutator quickly vanishes as Bob approaches the light cone and vanishes in the interior.

To emphasize the lesson learned in this section, unlike the (3+1) dimensional case, in (2+1) dimensions \textit{there is no such thing as timelike entanglement harvesting} at leading order in perturbation theory, as all entanglement obtained from the timelike region are all due to the field commutator.

\subsection{(1+1) dimensions}

\begin{figure*}[htp]
    \centering
    \includegraphics[scale=0.64]{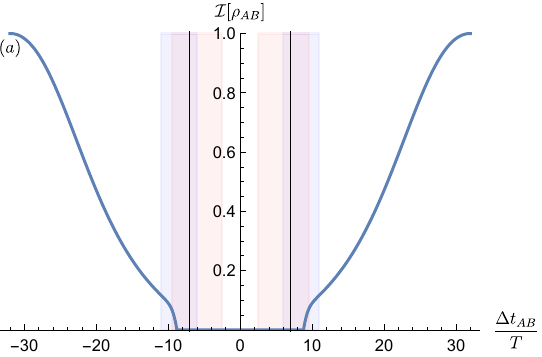}
    \includegraphics[scale=0.64]{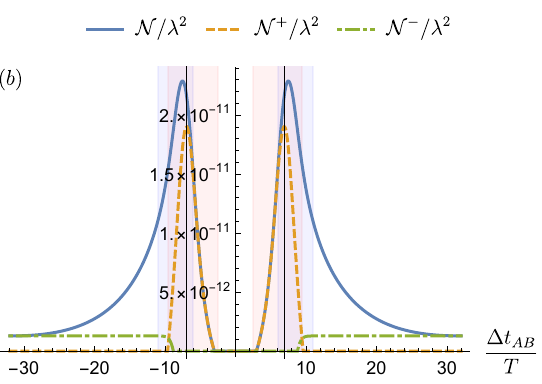}
    \includegraphics[scale=0.64]{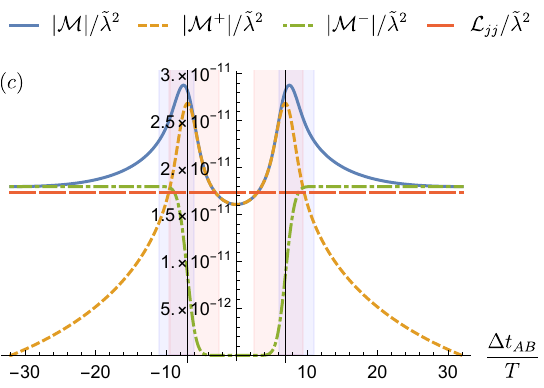}
    \caption{\textbf{Bipartite entanglement as a function of time delay} $t_\ab$ \textbf{between Alice and Bob's switching in (1+1) dimensions.} The parameters are $\Omega T = 7$ and $L=7T$. The infrared cutoff is chosen to be $\Lambda = 0.02 T^{-1}$. The vertical straight lines are the light cones of detector $A$ emanating from the event $(t_\textsc{a},\bm{0})$. The red shaded region marks the strong support of Alice's switching function, and the blue-shaded area marks the region where the behaviour of $|\mathcal{M}^\pm|$ starts to change dramatically.  (a) The communication-assisted entanglement estimator. (b) $\mathcal{N},\mathcal{N}^\pm$ as a function of $t_\ab$. Crucially, the anti-commutator part $|\mathcal{M}^+|$ vanishes in the interior of the light cone while the commutator part $|\mathcal{M}^-|$ increases, approaching a constant value. (c) $|\mathcal{M}|,|\mathcal{M}^\pm|,\mathcal{L}_{jj}$ as a function of $t_\ab$. }
    \label{fig: concurrence1D-1}
\end{figure*}

For completeness, we include the (1+1) dimensional case, that we show in Figure~\ref{fig: concurrence1D-1}. Note that due to the well-known infrared (IR) divergence for massless fields in (1+1)-dimensional Minkowski background, we do introduce a hard IR  cutoff $\Lambda T=0.02$ for the integrals over momentum as it is common in entanglement harvesting~(e.g., \cite{pozas2015harvesting}) and in other applications of quantum field theory~(e.g., \cite{birrell1984QFTCS}).

We see in Figure~\ref{fig: concurrence1D-1}(a) that the communication-based entanglement estimator continues to increase as Bob enters deep into the interior of Alice's light cone and eventually $\mathcal{I}[\hat{\rho}_\ab]\to 1$. This is because in (1+1) dimensions the field commutator is constant inside the light cone and thus the communication contribution $\mathcal{N}^-$ approaches a \textit{constant} value (see Figure~\ref{fig: concurrence1D-1}(b)), while the harvesting contribution $\mathcal{N}^+$ continues to decay as Bob goes deeper into the timelike interior of Alice's light cone. The constant nature of $\mathcal{N}^-$ is therefore a direct consequence of the constant commutator $C_1^-(\sx,\sx')$ in (1+1) dimensions (see Appendix~\ref{appendix: strong-Huygens})
\begin{align}
    C^-_1(\sx,\sx') &=  -\frac{\ii}{2}\text{sgn}(\Delta t)\Theta(|\Delta t| - |\Delta x|)\,.
\end{align}
Hence, similar to the (2+1)-dimensional case, there is no such as thing as timelike entanglement harvesting: the entanglement obtained by two timelike separated detectors mostly originates from field-mediated communication.

One subtlety concerning the (1+1)-dimensional massless scalar field is that IR cutoff plays a very significant role in influencing the results. More specifically, the commutator is independent of the IR cutoff but the anti-commutator is not. It is well-known that the Wightman function in (1+1) dimensions with a hard IR cutoff can be written as \cite{birrell1984QFTCS}
\begin{align}
    W(\sx,\sx') = -\frac{1}{4\pi}\log\left(-\Lambda^2\bigr[(\Delta t-\ii\epsilon)^2-\Delta x^2\bigr]\right)\,,
\end{align}
so the IR divergence contributes an additive constant that diverges in the limit $\Lambda\to 0$. Using \eqref{eq: commutator-anticomm}, it follows that  $C^-(\sx,\sx')$ is IR-safe because the additive constant drops out, but $C^+(\sx,\sx')$ has twice the additive constant. Consequently, the choice of IR cutoff will affect $\mathcal{M}^+$ and $\mathcal{L}_{jj}$ but not $\mathcal{M}^-$.

As a comparison, Figure~\ref{fig: concurrence1D-1} is plotted for $\Lambda T = 0.02$ while in \cite{pozas2015harvesting}, the calculation was done for $\Lambda T=0.001$. However, how much of the bipartite entanglement comes from harvesting depend on the IR cutoff chosen. Since both $|\mathcal{M}^+|$ and $\mathcal{L}_{jj}$ increases with decreasing the IR cutoff $\Lambda$ while $|\mathcal{M}^-|$ stays unchanged, spacelike entanglement harvesting improves as the cutoff $\Lambda$ decreases. We have checked that for $\Lambda T\lesssim 0.02$, we have $\mathcal{L}_{jj}>|\mathcal{M}^-|$, thus below this threshold the two detectors cannot get entangled even when Bob is in the interior of Alice's light cone despite the decay of the anti-commutator contribution $|\mathcal{M}^+|$. This is the reason why we chose slightly larger IR cutoff $\Lambda T = 0.02$, so that we can still see the impact of field-mediated communication on the bipartite entanglement.

Beyond the dependence on IR cutoff---which is a known pathology of the usual UDW model in free-space 1+1 dimensions--- the results obtained in Figure~\ref{fig: concurrence1D-1} parallels the one in (2+1) dimensions as they both have commutators that have nonzero support for timelike separated regions. Hence our main claim remains unchanged: as the detectors become causally connected, the field-mediated communication dominates the entanglement between the two detectors while the harvesting contribution diminishes.

\subsection{General comments on entanglement harvesting outside the UDW model in flat spacetime}

In this subsection, we are going to summarize some generic implications of our results based on massless scalar fields in  (1+1), (2+1) and (3+1) dimensions in Minkowski spacetime and then make some comments on more complicated spacetime backgrounds and different couplings.

The fact that in (3+1) dimensions the null-separated case is completely dominated by communication implies that one should be careful when deeming the entanglement obtained by the two detectors to  harvesting when they are null-connected. This includes, for instance, (1+1)-dimensional models involving derivative coupling variants of the Unruh-DeWitt model~\cite{Tjoa2020vaidya,Gallock-Yoshimura2021freefall} where the commutator of the field's proper time derivatives has support only along the null direction; setups involving massless fields conformally coupled to gravity in conformally flat backgrounds; or setups when one uses compactly supported switching but Alice and Bob's spatial smearings can be null-connected (e.g., some of the regimes in~\cite{henderson2018harvestingBH,Nadine2021delocharvesting}). Outside of conformal symmetry, one still needs to be careful as curvature can have non-trivial effects on the ability of null and timelike connected detectors to harvest entanglement.
{For example, in a black hole spacetime (such as Schwarzschild) it is possible to find scenarios where null communication through secondary geodesics allow for genuine entanglement harvesting~\cite{Casals2021commBH2}. In Kerr geometry, one cannot even find conformally flat slicing (unlike Schwarzschild geometry in Painlev\'e-Gullstrand coordinates \cite{Price2000kerrslice,DeFelice2019kerrslice}), thus the role of vacuum entanglement vs communication is likely to be even more complicated.}

The fact that timelike entanglement harvesting does not occur at all in (2+1) dimensions also implies that one should in general be very careful in ascribing the entanglement obtained by the two detectors to vacuum entanglement harvesting when the strong Huygens' principle does not hold. This includes, for instance, setups where the background geometry is curved and not maximally symmetric, such as cosmological spacetimes with minimal coupling; black hole spacetimes, including the lower-dimensional cases such as (rotating) Ba\~nados-Teitelboim-Zanelli (BTZ) black holes \cite{henderson2018harvestingBH,robbins2020entanglement}; and lower dimensional maximally symmetric spacetimes such as (2+1)-dimensional Anti-de Sitter geometry (AdS$_3$) \cite{henderson2019AdSharvesting}. Another relevant example involves a particular setup in (2+1) dimensions involving indefinite causal ordering (ICO). This was also recently investigated in \cite{Henderson2020temporal}, or superposition of trajectories \cite{Josh2021entanglement}. In light of our results, while there is not much doubt that there are quantum advantages due to ICO, when there is causal connection between the detectors one may wonder how much of this can be ascribed to enhancement of communication  (which is possible, see e.g. \cite{Ebler2018commICO}) or true enhancement of the vacuum harvesting protocol.

\section{Further results: Massive fields, higher dimensions and compact switchings}
\label{sec: further-results}

In this section we briefly discuss the the effect of the mass of the scalar field, the number of spacetime dimensions and the effect of using truly compact swtichings (instead of Gaussian ones) in light of the results obtained in the previous section. 


\subsection{Strong Huygens' principle in higher dimensions}

\begin{figure*}[tp]
    \centering
    \includegraphics[scale=0.64]{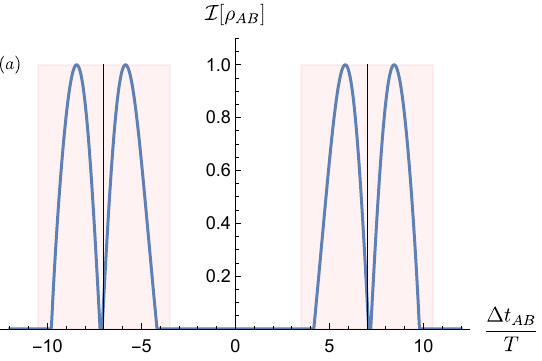}
    \includegraphics[scale=0.64]{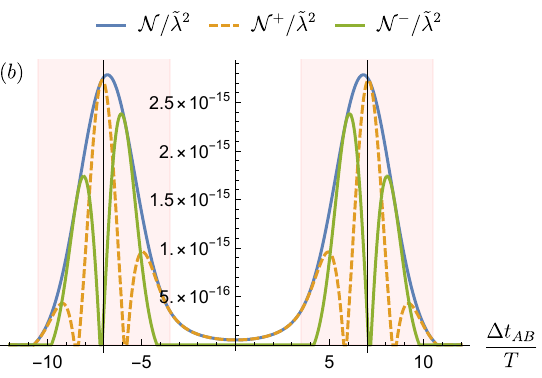}
    \includegraphics[scale=0.64]{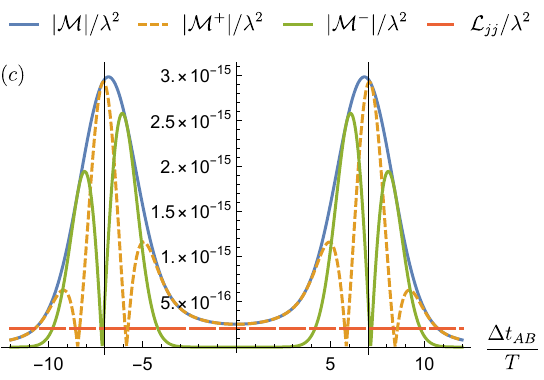}
    \caption{\textbf{Detector entanglement as a function of time delay} $t_\ab$ \textbf{between their switching peaks in (5+1) dimensions.} The parameters are $\Omega T = 7$ and $L=7T$. The vertical straight lines are the light cones of detector $A$ emanating from the event $(t_\textsc{a},\bm{0})$. The red-shaded region denotes Alice's light cone arising from the strong support $S_\textsc{a}$. Note the increasing number of peaks in all the plots compared to the $(3+1)$ dimensions.}
    \label{fig: concurrence5D-1}
\end{figure*}

As we briefly mentioned in Section~\ref{sec: setup}, when the strong Huygens' principle is satisfied, the field commutator $C^-(\sx,\sx')$ has support only along the null directions. For a Klein-Gordon field in $(n+1)$-dimensional Minkowski spacetimes, this occurs only when $n\geq 3$ is odd and for massless fields. It turns out that due to the structure of the commutator in higher dimensions, the role of communication manifests somewhat differently even if the principle is satisfied. A representative example is shown in Figure~\ref{fig: concurrence5D-1} for $n=5$.

Figure~\ref{fig: concurrence5D-1}(a) shows that like in the (3+1)-dimensional case, the communication-assisted entanglement estimator dominates at the neighbourhood of the light cone. However, notice that there are \textit{two peaks} around the light cone emanating from the centre of Alice's strong support, which suggests that while communication dominates in the neighbourhood of Alice's light cone (red shaded region), the \textit{anti-commutator} dominates around the region of maximum light-contact  $\Delta t_{\ab}=L$. This is because both $|\mathcal{M}^\pm|$ exhibit an extra peak, which leads to an additional peak in $\mathcal{N}^\pm$ in Figure~\ref{fig: concurrence5D-1}(b) and (c). Note that since the anti-commutator has three peaks around the light cone $\Delta t_\ab=L$, and the commutator only two peaks, for $n=5$ the commutator actually is not the dominant contribution at $\Delta t_\ab=L$, unlike for $n=3$. In fact, one can check that for odd  $n=2j+1$ with $j\geq 1$,  we have $j+1$ peaks for the anti-commutator around Alice's light cone and $j$ peaks for the commutator, thus the importance of the commutator at the light cone depends on whether $j$ is even or odd. {Note that we also see a similar asymmetry of $\mathcal{I}[\hat{\rho}_\ab]$ around the region of maximum light contact emanating from Alice's Gaussian peak at $\Delta t_\ab=7T$ as was the case in (3+1) dimensions.}

\begin{figure*}[tp]
    \centering
    \includegraphics[scale=0.64]{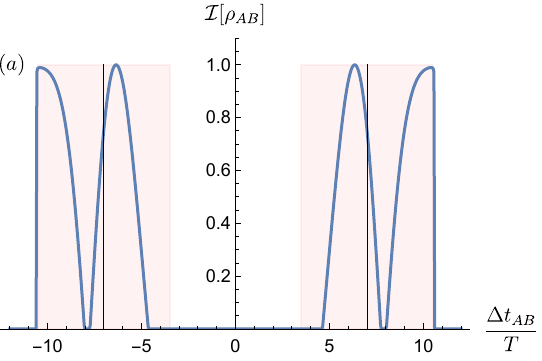}
    \includegraphics[scale=0.64]{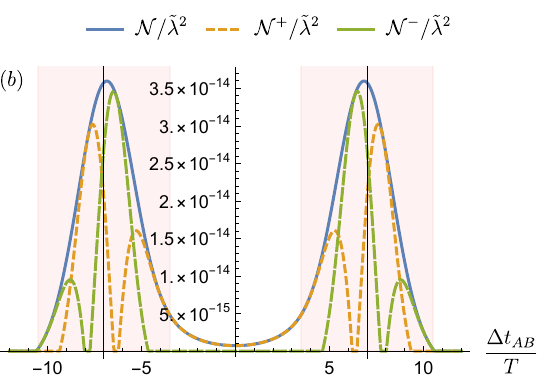}
    \includegraphics[scale=0.64]{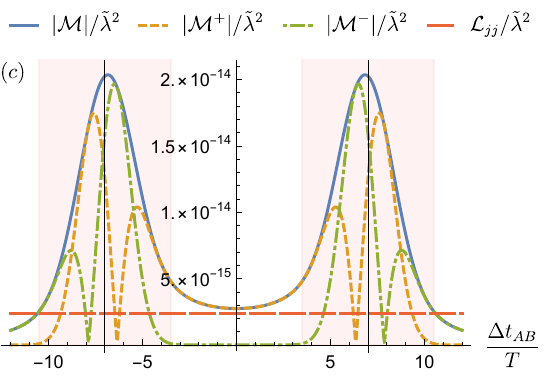}
    \caption{\textbf{Detector entanglement as a function of time delay} $t_\ab$ \textbf{between their switching peaks in (4+1) dimensions.} The parameters are $\Omega T = 7$ and $L=7T$. The vertical straight lines are the light cones of detector $A$ emanating from the event $(t_\textsc{a},\bm{0})$. The red-shaded region denotes Alice's light cone arising from the strong support $S_\textsc{a}$. Note the increasing number of peaks in all the plots compared to the $(2+1)$ dimensions.}
    \label{fig: concurrence4D-1}
\end{figure*}


The increasing number of peaks for both commutator and anti-commutator contributions can in fact be directly traced back to the behaviour of the imaginary and real parts of the Wightman function. For a massless field, the Wightman function for arbitrary $n$ reads (see, e.g., \cite{Gray2021shockwave,Takagi1986noise}, or by taking the small $m\to 0^+$ limit of massive scalar case in Appendix~\ref{appendix: Wightman-arbitrary}) 
\begin{align}
    W(\sx,\sx') &=  \frac{(-\ii)^{n-1}\Gamma(\frac{n-1}{2})}{4\pi^{\frac{n+1}{2}}[(\Delta t-\ii\epsilon)^2-|\Delta \bx|^2]^{\frac{n-1}{2}}}\,,
    \label{eq: Wightman-massless-D-dim-main}
\end{align}
where the $\epsilon$ is a UV regulator and the (distributional) limit $\epsilon\to 0$ is taken after integration:  for small $\epsilon>0$, the real and imaginary parts of~\eqref{eq: Wightman-massless-D-dim-main} corresponds to the ``nascent'' family whose limit $\epsilon\to 0$ is the Wightman function. The real and imaginary part in that distributional limit yield respectively the (vacuum expectation of) the anti-commutator and the commutator.

The simple case of the commutator can be actually computed easily from a mode expansion (see Appendix~\ref{appendix: strong-Huygens}). For arbitrary odd $n\geq 3$ the (state independent) expectation of the commutator takes the form
\begin{align}
    C^-_{n}(\sx,\sx') &= \ii\sum_{j=0}^{\frac{n-3}{2}}\frac{a_j}{|\Delta \bx|^{n-2-j}}\bigg[\delta^{(j)}(\Delta t+|\Delta \bx|) \notag\\
    &\hspace{1.5cm} +(-1)^{j+1}\delta^{(j)}(\Delta t-|\Delta \bx|)\bigg]\,,
    \label{eq: commutator-n-dim}
\end{align}
where $a_j$ are real, $\Delta t=t-t'$, $\Delta \bx = \bx-\bx'$ and $\delta^{(j)}(z)$ is the $j$-th {distributional derivative} of the Dirac delta function. The distributional derivatives of Dirac deltas have support strictly along the null direction, but they differ from the Dirac delta in that the ``nascent'' family defining $\delta^{(j)}(z)$ has $j+1$ peaks\footnote{One can readily see this by using Gaussian functions as a family of nascent delta functions, and their derivatives define a family of derivatives of delta functions.}. Since the commutator is dominated by the highest derivative of the Dirac delta (the $(n-3)/2$-th derivative) for sufficiently large detector separations (which is the case in this work), the number of peaks in $|\mathcal{M}^-|$ is $1+(n-3)/2$. Thus for $n=5$, the highest derivative is $j=1$, which gives two peaks for the commutator contribution, in agreement with Figure~\ref{fig: concurrence5D-1}(c). It is straightforward to check that for $n=7$, we will have three peaks in $\mathcal{I}[\hat{\rho}_\ab]$ which follows from the number of peaks in $|\mathcal{M}^-|$, and this pattern continues to higher dimensions.

Similarly, there are also increasing number of peaks in $|\mathcal{M}^\pm|$ for even $n$. As shown in Figure~\ref{fig: concurrence4D-1}, we plot the case for $n=4$ and we see that  we also have more peaks in $|\mathcal{M}^\pm|$ (hence $\mathcal{N}^\pm$ and $\skri[\hat{\rho}_\ab]$) as compared to the $n=2$ case in Figure~\ref{fig: concurrence2D-1}. However, the pattern differs slightly from the odd $n$ case. More generally, for even $n=2\ell$ with $\ell \geq 1$ there will be $\ell$ peaks for both the anti-commutator and the commutator around Alice's light cone. Since the number of peaks around both components are equal, it is always the case for even $n$ that both components contribute equally to the bipartite entanglement around the light cone. Despite this we want to re-emphasize that for timelike contact entanglement is still dominated by communication in all even spatial dimensions, rather than true harvesting.

\subsection{Massive scalar field}

In this subsection we will obtain  analogous results for massive scalar fields.

\begin{figure*}[tp]
    \centering
    \includegraphics[scale=0.64]{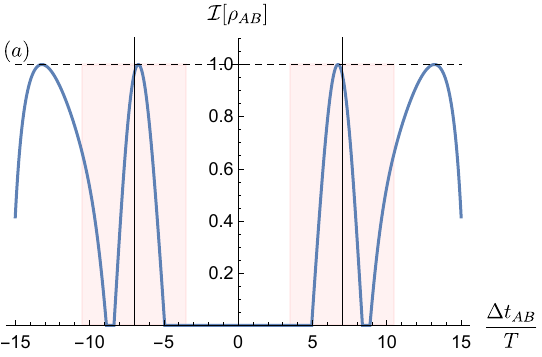}
    \includegraphics[scale=0.64]{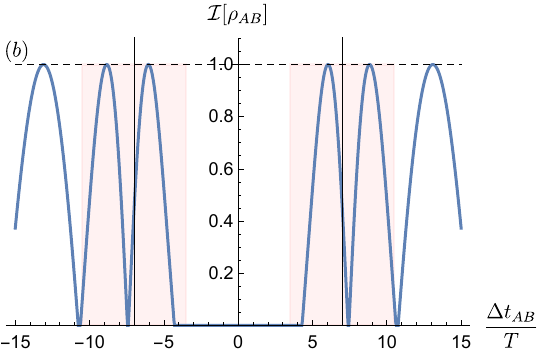}
    \includegraphics[scale=0.64]{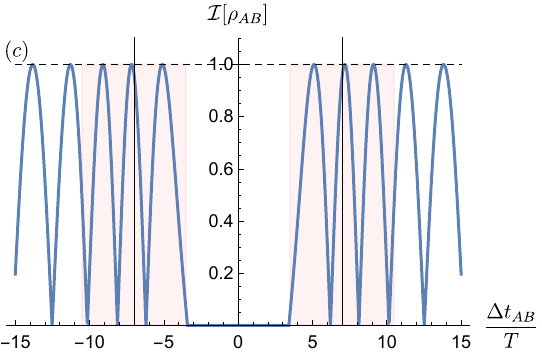}
    \includegraphics[scale=0.64]{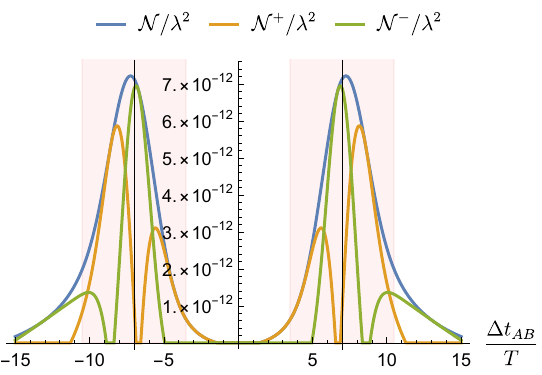}
    \includegraphics[scale=0.64]{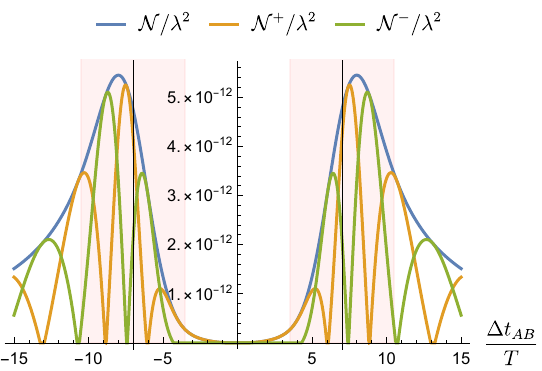}
    \includegraphics[scale=0.64]{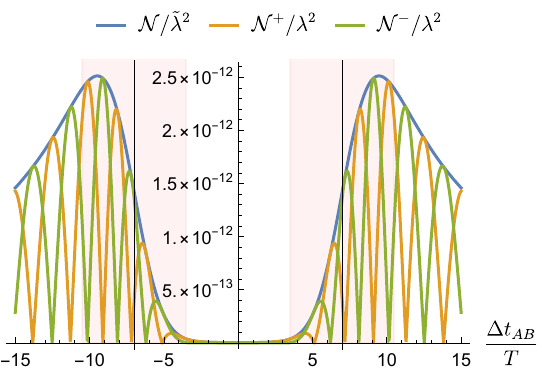}
    \caption{\textbf{Detector entanglement as a function of time delay} $t_\ab$ \textbf{between their switching peaks in (3+1) dimensions for massive scalar fields.}  (a) $mT = 0.2$ (b) $mT = 0.5$ (c) $mT = 1$. The parameters are $\Omega T = 7$ and $L=7T$. The vertical straight lines are the light cones of detector $A$ emanating from the event $(t_\textsc{a},\bm{0})$. The red-shaded region denotes Alice's light cone arising from the strong support $S_\textsc{a}$. Observe that for small mass the behaviour is close to massless fields and there increasing oscillatory behaviour as the mass of the field increases.}
    \label{fig: concurrence3D-massive}
\end{figure*}

We plot the massive field results in Figure~\ref{fig: concurrence3D-massive}. There are several important distinctive features as compared to the massless case. The first observation is that for a massive field the commutator has support inside the light cone regardless of the dimension of spacetime, even within the deep interior of Alice's light cone ($\Delta t_\ab\gg 7 T$). The second observation is that the oscillatory nature of both the commutator and anti-commutator contributions to the correlation term $|\mathcal{M}^\pm|$  become more pronounced as the mass of the field increases. The third observation is that the oscillations are not ``in phase'': the dominant contributions to entanglement alternate between the anti-commutator contribution and the commutator contribution, so that on average they both contribute equally for timelike separated detectors that are switched on long enough.

The oscillatory nature of both contributions can also be directly traced back to the behaviour of the imaginary and real parts of the Wightman function, which is given for arbitrary $m$ and $n$ by (see Appendix~\ref{appendix: Wightman-arbitrary} for derivation) 
\begin{align}
    W(\sx,\sx') &= \frac{m ^{\frac{n-1}{2}}}{(2\pi)^{\frac{n+1}{2}}}\frac{1}{[-({\Delta t-\ii\epsilon })^2+|\Delta\bx|^2]^{\frac{n-1}{4}}}\notag\\
    &\hspace{0.5cm}\times K_{\frac{n-1}{2}}(m\sqrt{-({\Delta t-\ii\epsilon})^2+|\Delta\bx|^2}) \,.
    \label{eq: Wightman-massive-D-dim-main}
\end{align}
We can regard the UV regulator $\epsilon$ as providing a nascent family of complex-valued functions whose limit gives the Wightman function above. By plotting the nascent family for finite nonzero $\epsilon$, one can see the same oscillatory behaviour of $|\mathcal{M}^\pm|$, including the number of peaks that appear in them.

Notice that while both massless fields in even spatial dimensions and massive fields  have commutators with support for timelike separation, their relative contribution to the entanglement generated between two timelike separated detectors are quite different.  Namely, on one hand for the massless case entanglement deep into the region of timelike separation is dominated by the commutator contribution and therefore it cannot be attributed to genuine harvesting. On the other hand, for the massive case both communication and harvesting can be thought of as contributing equally to the detectors' entanglement.



\subsection{Compactly supported switching function}
\label{subsec: compact-switching}

Finally, we complete our analysis by showing that the main claims of this work are not affected by the use of non-compact switching, as long as the strong supports of both detectors are in spacelike separation. We do this by performing the same calculations for compactly supported switching functions and restricting our attention to the simple case of a massless scalar field in (3+1) dimensions. Unlike the Gaussian case, there is not much in the way of simplification that we can effect for the matrix elements of $\hat{\rho}_\ab$, thus we calculate the matrix elements for the case of compact switchings numerically from
~\eqref{eq: Lij} and~\eqref{eq: M-nonloc} using the standard formula for the Wightman function for massless field in (3+1) dimensions using Eq.~\eqref{eq: Wightman-massless-D-dim-main}:
\begin{align}
    W(\sx,\sx') = -\frac{1}{4\pi^2}\frac{1}{(\Delta t-\ii\epsilon)^2-|\Delta \bx|^2}\,.
\end{align}

\begin{figure*}[tp]
    \centering
    \includegraphics[scale=0.8]{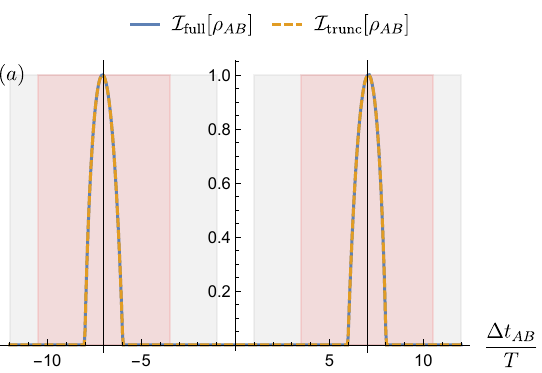}
    \includegraphics[scale=0.8]{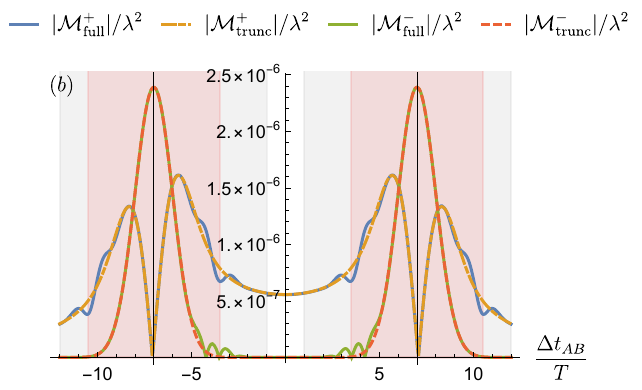}
    \caption{\textbf{Comparison between compact and non-compact switching on detector entanglement in (3+1) dimensions for massless scalar fields.} The parameters are $\Omega T = 4$ and $L=7T$. (a) Comparison of the communication-assisted entanglement estimator for full and truncated Gaussian switchings. (b) Comparison of $|\mathcal{M}^\pm|$ for the full and truncated Gaussian switchings. The truncated Gaussian has compact support $R_j = [-3T+t_j,3T+t_j]$. The vertical straight lines are the light cones of detector $A$ emanating from the event $(t_\textsc{a},\bm{0})$. The white region between the grey zones near the origin is the values of $t_\ab$ where the two compactly supported detectors can be truly spacelike separated. The red regions are Alice's light cone with respect to the full Gaussian switching's strong support.}
    \label{fig: concurrence3D-cut}
\end{figure*}

The compact switching we consider is the truncated Gaussian,
\begin{align}
    \chi_j^{\text{trunc}}(t)&= e^{-\frac{(t-t_j)^2}{T^2}}\Phi_{R_j}\,,
\end{align}
where $\Phi_{R_j}$ is the indicator function on the compact interval $R_j=[-3T+t_j,3T+t_j]$, given by
\begin{align}
    \Phi_{R_j} \coloneqq \begin{cases}
    1 \hspace{0.5cm}& t\in R_j\\
    0 & t\neq R_j\end{cases}\,.
\end{align}
This choice of truncated Gaussian allows us to compare the result with the full Gaussian switching more easily. As the detector separation is set at $L=7T$, in this case the two detectors can be made strictly spacetike separated without any tails putting them in marginal light contact. The comparison is shown in Figure~\ref{fig: concurrence3D-cut}. The grey shaded region marks the light cone of Alice's compact support if the switching is the truncated Gaussian, which spans interval of $6T$. The red shaded region marks the light cone of Alice's strong support if the switching is Gaussian.

Our example here gives essentially identical communication-assisted entanglement estimator $\mathcal{I}[\hat{\rho}_\ab]$ in Figure~\ref{fig: concurrence3D-cut}(a). We also see from Figure~\ref{fig: concurrence3D-cut}(b) up to small oscillations near the boundary of compact support, the use of compactly supported switching leads to essentially the same result as the non-compact switching: namely, the communication component (commutator contribution) dominates near the light cone while the vacuum harvesting component (anti-commutator contribution) vanishes. This is not surprising because the essential reason for the dominance of communication over harvesting at null separation is not influenced by the shape of the switching function but rather the distributional behaviour of the real and imaginary parts of the Wightman function\footnote{Unfortunately, at the time of writing our \textit{Mathematica} code was insufficient for obtaining results with nonzero \textit{spacelike} entanglement harvesting, so we picked $\Omega T=4$ where at least the $\mathcal{M}^\pm$ terms can be compared for the null separation. We are grateful to Patricia Ribes Metidieri and Sergi Nadal for using a different numerical code to show that they agree with our main claims for the spacelike harvesting regime with compact switching.}.

\section{Conclusions}

In this paper we analyzed whether entanglement harvesting can be achieved when particle detectors are causally connected and are able to exchange information and therefore get entangled without harvesting correlations from the field.

In particular, we studied the role of the field-mediated communication in the so-called entanglement harvesting protocol for the Minkowski vacuum in arbitrary spacetime dimensions. By varying the time delay between the switching functions of two detectors and hence their causal relationships, we investigated how much of the entanglement acquired by the two detectors after interaction with the fields is due to field-mediated communication between them and how much is due to vacuum entanglement harvesting.

More specifically, the communication between two detectors at leading order communication between the two detectors is mediated by the field commutator (see e.g.,~\cite{Jonsson2014cavityQED,Causality2015Eduardo,Casals2020commBH}), which does not care about the correlations pre-existing in the field since it is state independent. Therefore its contribution would be the same whether the field state has correlations to harvest or not. We hence argue that the ability to harvest entanglement is mediated by the field anti-commutator. Both the commutator and the anti-commutator have very different behaviour depending on the dimensions of spacetime.

We compared the contribution of the commutator and the anti-commutator to the entanglement acquired by two detectors interacting with the field. We showed that, for massless fields in any dimensions, when the two detectors are causally connected, the entanglement they acquire does not come from harvesting. Instead, it is dominated by the field-state independent commutator contribution to the correlation between the detectors, hence being due to communication and not harvesting as has been sometimes claimed. 

We have also analyzed the case of massive fields, where the behaviour is somewhat different: for massive enough fields the contributions of harvesting and communication to the entanglement acquired by the detectors in causal connection tends to be equally contributed by both communication and harvesting. 

Finally we have considered how our results also apply to more general scenarios such as curved spacetimes, smeared detectors, detectors with indefinite causal order or derivative coupling UDW models. In this context, we have discussed how the results about harvesting vs communication largely apply to all these general scenarios.

The key takeaway in view of our results is that for a genuine ``entanglement harvesting protocol'',  the entanglement `swapped' from the existing field correlations should be the major contributor to the bipartite entanglement between the detectors. In this context, we have seen that in the cases when the field commutator is the leading contribution to the entanglement between detectors, the entanglement cannot come from harvested correlations. This is so because the commutator contribution is the same regardless of the state of the field and hence it will entangle the detectors in the same way whether the field has pre-existing correlations or not. That is the case in most massless field scenarios when the two detectors are in causal contact, where their entanglement comes from their ability to signal each other via the field. Our results emphasize the importance of remaining spacelike separated to properly claim that the detectors harvest entanglement from the field.

\section*{Acknowledgement}
E. T. acknowledges generous support from the Mike and Ophelia Lazaridis Fellowship. E. M-M. acknowledges support through the Natural Sciences and Engineering Research Council of Canada (NSERC) Discovery Grants program as well as Ontario Early Researcher Award. This work is conducted on the traditional territory of the Neutral, Anishnaabeg, and Haudenosaunee Peoples. The University of Waterloo and the Institute for Quantum Computing are situated on the Haldimand Tract, land that was promised to Six Nations, which includes six miles on each side of the Grand River.

\appendix

\onecolumngrid 

\section{Non-local term in arbitrary dimensions}
\label{appendix: M-proof}

Here we will derive the non-local term $\mathcal{M}$, generalizing the result of \cite{pozas2015harvesting} to arbitrary number of spatial dimensions $n$ and mass $m$. We will also show how the derivation of  $\mathcal{M}$ conveniently splits the contributions coming from the field commutator and anti-commutator.

First, we recall that we have two identical detectors which are pointlike and at rest relative to the quantization frame with Minkowski coordinates $(t,\bx)$. The detector trajectories $\sx_j(t)$  ($j=\textsc{a,b}$) are static relative to the quantization frame so we can write $\sx_{j}(t) = (t,\bx_j)$ where $\bx_j$ are constant. The detectors are turned on for the same effective duration (controlled by Gaussian width $T$) but they are allowed to be turned on at different times (different Gaussian peaks in Eq.~\eqref{eq: Gaussian-switch}). We will comment on the inclusion of spatial smearing at the end of this section. 

Under these assumptions, the non-local contribution $\mathcal{M}$ can be written as
\begin{align}
    \mathcal{M} = -\lambda^2\int_{-\infty}^\infty\!\!\!\dd t\int^t_{-\infty}\!\!\!\dd t'e^{\ii\Omega(t+t')}\bigg[&e^{-\frac{(t-t_\textsc{a})^2}{T^2}} e^{-\frac{(t'-t_\textsc{b})^2}{T^2}}\int \frac{\dd^n\bk}{2(2\pi)^n\omega_\bk}e^{-\ii\omega_\bk (t-t')+\ii\bk\cdot(\bx_\textsc{a}-\bx_\textsc{b})}+(\textsc{a}\leftrightarrow \textsc{b})\bigg]\,,
\end{align}
where we have implemented the time ordering $\Theta(t-t')$ as a nested integral and $\omega_\bk=\sqrt{|\bk|^2+m^2}$ is the relativistic dispersion relation. It is convenient to perform the following redefinition and change of variables:
\begin{align}
    t_{\ab} \coloneqq t_{\textsc{b}}-t_\textsc{a}\,,\hspace{0.5cm}\bx_{\ab} \coloneqq \bx_{\textsc{b}}-\bx_\textsc{a}\,,\hspace{0.5cm}  t\to t-t_\textsc{a}\,,\hspace{0.5cm} t'\to  t'-t_\textsc{a}\,.
\end{align}
This will give a more symmetric expression
\begin{align}
    \mathcal{M} = -\lambda^2\int_{-\infty}^\infty\!\!\!\dd t\int^t_{-\infty}\!\!\!\dd t'\int \frac{\dd^n\bk}{2(2\pi)^n\omega_\bk}e^{-\ii\omega_\bk (t-t')} e^{\ii\Omega(t+t'+{2t_\textsc{a}})}\bigg[&e^{-\frac{t^2}{T^2}} e^{-\frac{(t'- t_\ab)^2}{T^2}}e^{-\ii\bk\cdot \bx_\ab}+e^{-\frac{(t-t_\ab)^2}{T^2}} e^{-\frac{{t'}^2}{T^2}}e^{\ii\bk\cdot\bx_\ab}\bigg]\,.
\end{align}

Let us rewrite this in a more compact form
\begin{align}
    \mathcal{M} &= -\lambda^2e^{2\ii\Omega t_\textsc{a}}\int \frac{\dd^n\bk}{2(2\pi)^n\omega_\bk}\mathcal{K}(\bk)\,,\label{eq: full-M}\\
    \mathcal{K}(\bk) &\coloneqq \int_{-\infty}^\infty\!\!\!\dd t\int^t_{-\infty}\!\!\!\dd t'e^{-\ii\omega_\bk (t-t')} e^{\ii\Omega(t+t')}\bigg[e^{-\frac{t^2}{T^2}} e^{-\frac{(t'- t_\ab)^2}{T^2}}e^{-\ii\bk\cdot \bx_\ab}+e^{-\frac{(t-t_\ab)^2}{T^2}} e^{-\frac{{t'}^2}{T^2}}e^{\ii\bk\cdot\bx_\ab}\bigg]\,,
\end{align}
where we keep all the global phases for clarity. The integral can be done in closed form:
\begin{align}
    \mathcal{K}(\bk) &= \frac{\pi}{2}T^2e^{-\frac{T^2}{2}\rr{\Omega^2+\omega_\bk^2}}\bigg[e^{\ii\bk\cdot\bx_{\ab}+\ii t_{\ab}(\Omega-\omega_\bk)}+e^{-\ii\bk\cdot\bx_{\ab}+\ii t_{\ab}(\Omega+\omega_\bk)}\bigg]\notag\\
    &+ \frac{\sqrt{\pi}}{2}T^2e^{-\frac{t_\ab^2}{T^2}}\bigg[e^{\ii\bk\cdot\bx_{\ab}}\mathcal{J}\rr{\frac{T(\omega_\bk+\Omega)}{2},T(\omega_\bk-\Omega)+\frac{2\ii t_{\ab}}{T}}+e^{-\ii\bk\cdot\bx_{\ab}}\mathcal{J}\rr{\frac{T(\omega_\bk+\Omega)}{2}-\frac{\ii t_{\ab}}{T},T(\omega_\bk-\Omega)}
    \bigg]\,,
    \label{eq: scriptK}
\end{align}
where we define 
\begin{align}
    \mathcal{J}(a,b)\coloneqq -\ii\sqrt{\pi}e^{-a^2-\frac{b^2}{4}}\text{erfi}\rr{\frac{a+b/2}{\sqrt{2}}}\,,
\end{align}
with $\text{erfi}(z) = -\ii\, \text{erf}(\ii z)$ and $\text{erf}(z)$ is the error function.

Next, we separate the radial and angular part of the integration measure in \eqref{eq: full-M}:
\begin{align}
    \int\frac{\dd^n\bk}{2(2\pi)^n{\omega_\bk}}&= \frac{1}{2(2\pi)^n}\int_0^\infty\dd|\bk| {\frac{|\bk|^{n-1}}{\sqrt{|\bk|^2+m^2}}} \int\dd\Omega_{n-1} = \frac{1}{2(2\pi)^n}\int_0^\infty\dd|\bk| {\frac{|\bk|^{n-1}}{\sqrt{|\bk|^2+m^2}}}\int \dd\mu_{n-2}\int_0^\pi\dd\theta \sin^{n-2}\theta\,,
\end{align}
where $\dd\Omega_{n-1}$ is the area element of the unit sphere $S^{n-1}$ and $\dd\mu_{n-2}$ is the remaining angular part of the integration measure:
\begin{align}
    \dd\Omega_{n-1} &= \dd\theta (\sin\theta)^{n-2}\dd\mu_{n-2}\,,\hspace{0.5cm}
    \dd\mu_{n-2} \coloneqq\prod_{i=1}^{n-2}\dd\varphi_i\,(\sin\varphi_i)^{n-2-i}\,.
\end{align}

The integral over $\dd\mu_{n-2}$ can be found using the trick in~\cite{tjoa2021makes} as follows: 
\begin{align}
    \int\dd\Omega_{n-1}  &= 
    \int \dd \mu_{n-2}\int_0^\pi\dd\theta \,\sin^{n-2}\theta= \frac{2\pi^{\frac{n}{2}}}{\Gamma(\frac{n}{2})}\,,\\
    \int \dd\theta\sin^{n-2}\theta &= \frac{\sqrt{\pi}\Gamma(\frac{n-1}{2})}{\Gamma(\frac{n}{2})} \Longrightarrow \int\dd\mu_{n-2} = \frac{2 \pi ^{\frac{n-1}{2}}}{\Gamma \left(\frac{n-1}{2}\right)}\,.
\end{align}
Hence we get
\begin{align}
    \int\frac{\dd^n\bk}{2(2\pi)^n{\omega_\bk}}&= \frac{1}{2(2\pi)^n}\frac{2 \pi ^{\frac{n-1}{2}}}{\Gamma \left(\frac{n-1}{2}\right)}\int_0^\infty\dd|\bk| {\frac{|\bk|^{n-1}}{\sqrt{|\bk|^2+m^2}}} \int_0^\pi\dd\theta \sin^{n-2}\theta\,.
    \label{eq: angular-integral}
\end{align}
The only component that depends on the angular variable is the phase $e^{\pm \ii\bk\cdot\bx_{\ab}}$, thus we can perform this integral first:
\begin{align}
    \int_0^\pi\dd\theta \sin^{n-2}\theta\,e^{\pm\ii\bk\cdot\bx_{\ab}} 
    &= \int_0^\pi\dd\theta \sin^{n-2}\theta\,e^{\pm\ii|\bk||\bx_{\ab}|\cos\theta} 
    = \sqrt{\pi } \Gamma\left(\frac{n-1}{2}\right) \, {_0\tilde{F}_1}\left(\frac{n}{2};-\frac{|\bk|^2|\bx_\ab|^2}{4}\right)\,,
    \label{eq: angular-integral-final}
\end{align}
where ${_0\tilde{F}_1}$ is the regularized generalized hypergeometric function \cite{NIST:DLMF}. For completeness, we note that this could also be equivalently written in terms of Bessel function using the fact that for $n>1$ we have \cite{abramowitz1965handbook}
\begin{align}
     {_0\tilde{F}_1}\left(\frac{n}{2};-\frac{|\bk|^2|\bx_\ab|^2}{4}\right) = \left(\frac{2}{|\bk||\bx_\ab|}\right)^{\frac{n-2}{2}}J_{\frac{n-2}{2}}(|\bk||\bx_\ab|)\,.
     \label{eq: bessel-angle}
\end{align}
This is also called the Bessel-Clifford function, denoted as $\mathcal{C}_n(z)={_0\tilde{F}_1}(n+1;z)$ \cite{abramowitz1965handbook}.

Since $\mathcal{K}(\bk)$ in \eqref{eq: scriptK} has four terms, it is convenient to rewrite the expression as $\mathcal{K} = \mathcal{K}_1+ \mathcal{K}_2 + \mathcal{K}_3 + \mathcal{K}_4$, where
\begin{align}
    \mathcal{K}_1(|\bk|) &= 2^{-n-1} \pi ^{1-\frac{n}{2}} T^2 \, _0\tilde{F}_1\left(\frac{n}{2};-\frac{|\bk|^2|\bx_\ab|^2}{4} \right) e^{-\frac{1}{2} T^2 \left(\tcr{\omega_\bk^2+\Omega ^2}\right)+ \ii t_{\ab}  \tcr{(\Omega -\omega_\bk)}}\,,\\
    \mathcal{K}_2(|\bk|) &= 2^{-n-1} \pi ^{1-\frac{n}{2}} T^2 \, _0\tilde{F}_1\left(\frac{n}{2};-\frac{|\bk|^2|\bx_\ab|^2}{4}\right) e^{-\frac{1}{2} T^2 \left(\tcr{\omega_\bk^2+\Omega ^2}\right)+ \ii t_{\ab}  \tcr{(\Omega+\omega_\bk )}}\,,\\
    \mathcal{K}_3(|\bk|) &= -\ii 2^{-n} \pi ^{\frac{1}{2}-\frac{n}{2}} T^2 e^{\ii t_{\ab}  \Omega -\frac{t_{\ab} ^2}{2 T^2}-\frac{T^2 \Omega ^2}{2}} \mathcal{F}\left(\frac{\tcr{\omega_\bk} T^2+\ii t_{\ab} }{\sqrt{2} T}\right) \, _0\tilde{F}_1\left(\frac{n}{2};-\frac{|\bk|^2|\bx_\ab|^2}{4}\right) \,,\\
    \mathcal{K}_4(|\bk|) &= -\ii 2^{-n} \pi ^{\frac{1}{2}-\frac{n}{2}} T^2 e^{\ii t_{\ab}  \Omega -\frac{t_{\ab} ^2}{2 T^2}-\frac{T^2 \Omega ^2}{2}} \mathcal{F}\left(\frac{\tcr{\omega_\bk}T^2-\ii t_{\ab} }{\sqrt{2} T}\right) \, _0\tilde{F}_1\left(\frac{n}{2};-\frac{|\bk|^2|\bx_\ab|^2}{4} \right)\,,
\end{align}
where $\mathcal{F}(z)\coloneqq e^{-z^2}\int_0^z\dd y\, e^{y^2}$ is the Dawson's integral \cite{NIST:DLMF}. We have made explicit the fact that $\mathcal{K}_j$ depends only on the magnitude of the momentum vector $|\bk|$ \tcr{(since $\omega_\bk=\sqrt{|\bk|^2+m^2}$)}, thus it is convenient to write $k\coloneqq |\bk|$. The full expression for $\mathcal{M}$ now reads
\begin{align}
    \mathcal{M} &= -\lambda^2e^{2\ii\Omega t_\textsc{a}}\int_0^\infty\!\!\! \dd k\,{\frac{k^{n-1}}{\sqrt{k^2+m^2}}}\sum_{j=1}^4 \mathcal{K}_j(k)\,.
    \label{eq: full-M-final}
\end{align}
This is the final expression for the non-local matrix element $\mathcal{M}$ for arbitrary mass $m\geq 0$.

In what follows we would like to be able to split $\mathcal{M}$ into two parts, one which depends only on the anti-commutator, denoted by $\mathcal{M}^+$, and the other which depends only on the field commutator, denoted by $\mathcal{M}^-$. This split is necessary for splitting harvesting contribution (which depends on anti-commutator) from communication contribution (which depends on commutator). Let us write the expectations of the the anti-commutator and the commutator in terms of the Wightman function:
\begin{align}
    C^\pm(\sx,\sx') &= W(\sx,\sx')\pm W(\sx',\sx)\,.
\end{align}
Remarkably, what is perhaps not obvious from the splitting of $\mathcal{M}$ into $\mathcal{K}_j$'s is that the field anti-commutator expectation  $C^+(\sx,\sx')$ depends only on $\mathcal{K}_1$ and $\mathcal{K}_2$, while the field commutator expectation $C^-(\sx,\sx')$ depends only on  $\mathcal{K}_3$ and $\mathcal{K}_4$. Consequently, the (anti-)commutator contributions can be written as $ \mathcal{M}^\pm = \mathcal{M} \pm \mathcal{M}'$, where $\mathcal{M}'$ is the same integral as $\mathcal{M}$ in Eq.~\eqref{eq: full-M} but with the replacement $\omega_\bk\to-\omega_\bk$ and $\bk\to-\bk$. Under these replacements, we have
\begin{align}
    \mathcal{M}' &= -\lambda^2e^{2\ii\Omega t_\textsc{a}}\int_0^\infty\!\!\! \dd k \sum_{j=1}^4 \mathcal{K}'_j(k)\,,
\end{align}
where as before we use $k=|\bk|$ and
\begin{align}
    \mathcal{K}_1'(k) &= \mathcal{K}_2(k)\,,\hspace{0.5cm}
    \mathcal{K}_2'(k) = \mathcal{K}_1(k)\,,\hspace{0.5cm}
    \mathcal{K}_3'(k) = -\mathcal{K}_4(k)\,,\hspace{0.5cm}
    \mathcal{K}_4'(k) = -\mathcal{K}_3(k)\,.
\end{align}
Hence, the (anti-)commutator contributions to $\mathcal{M}$ are compactly expressible as
\begin{align}
    \mathcal{M}^+ &= -\lambda^2e^{2\ii\Omega t_\textsc{a}}\int_0^\infty\!\!\! \dd k\,{\frac{k^{n-1}}{\sqrt{k^2+m^2}}}\rr{\mathcal{K}_1(k)+\mathcal{K}_2(k)}\,,\\
    \mathcal{M}^- &= -\lambda^2e^{2\ii\Omega t_\textsc{a}}\int_0^\infty\!\!\! \dd k\,{\frac{k^{n-1}}{\sqrt{k^2+m^2}}}\rr{\mathcal{K}_3(k)+\mathcal{K}_4(k)}\,.
\end{align}

\section{Spatially smeared detector}
\label{appendix: spatial smearing}

The calculation for the case of a spatially smeared detector is straightforward. For simplicity, we will consider the special case where both detectors have identical spatial smearing and switching functions (up to spacetime translation), with the same inertial trajectory at rest in the quantization frame.

Under these assumptions, we can write $\chi_j(t)\coloneqq \chi(t-t_j)$  where $j=\text{A}, \text{B}$ and $\chi(t)$ is some real function. The spatial smearing of both detectors is a common real-valued function $F(\bx)$ that is $L^1$-normalized to unity and we write $F_j(\bx) = F(\bx-\bx_j)$.  The resulting matrix elements in \eqref{eq: Lij} and \eqref{eq: M-nonloc} are modified into
\begin{align}
    \mathcal{L}_{ij} &= \lambda^2\int \dd t\,\dd t'\int\dd^n\bx\,\dd^n\bx'\,\chi_i(t)\chi_j(t')F_i(\bx)F_j(\bx') e^{-\ii\Omega(t - t')}W(t,{\bx} ;t',\bx')
    \label{eq: Lij-smeared}\\
    \mathcal{M} &= -\lambda^2\int \dd t\,\dd t'\int\dd^n\bx\,\dd^n\bx'\, e^{\ii\Omega(t+t')} \chi_\textsc{a}(t)\chi_\textsc{b}(t')F_\textsc{a}(\bx)F_\textsc{b}(\bx') \Theta(t-t') W(t,\bx;t',\bx')\notag\\
    &\hspace{3.8cm}+ e^{\ii\Omega(t+t')}\chi_\textsc{a}(t)\chi_\textsc{b}(t')F_\textsc{a}(\bx)F_\textsc{b}(\bx')\Theta(t' - t)W(t',\bx';t,\bx)\bigr]\,.
    \label{eq: M-nonloc-smeared}
\end{align}
The expression for $\mathcal{L}_{ij}$ in Eq.~\eqref{eq: Lij} is modified to
\begin{align}
    \mathcal{L}_{ij} = \lambda^2\int \frac{\dd^n\bk}{2(2\pi)^n\omega_\bk}\tilde{\chi}_i(\Omega+\omega_\bk)\tilde{\chi}^*_j(\Omega+\omega_\bk)\tilde F_i(\bk)\tilde F_j^*(\bk)\,,
\end{align}
where $\tilde F_i(\bk)$ is the Fourier transform of $F_i(\bx)$. The translation property of the Fourier transform allow us to write this as
\begin{align}
    \mathcal{L}_{ij} = \lambda^2\int \frac{\dd^n\bk}{2(2\pi)^n\omega_\bk}|\tilde{\chi}(\Omega+\omega_\bk)|^2|\tilde F(\bk)|^2e^{-\ii(\Omega+\omega_\bk)(t_i-t_j)}e^{\ii\bk\cdot (\bx_i-\bx_j)}\,.
\end{align}
Note that $\mathcal{L}_{jj}$ (the excitation probability of detector $j$) is independent of $t_j$ and $\bx_j$, as we expect from translational invariance. The pointlike limit is recovered simply by setting $\tilde{F}(\bk)=1$. 

For the non-local $\mathcal{M}$ matrix element we can proceed similarly.  For the Gaussian switching considered in Appendix~\ref{appendix: M-proof}, the resulting expression for $\mathcal{M}$ in \eqref{eq: full-M} turns out to be obtainable by simply replacing
\begin{align}
    \mathcal{K}(\bk)\to |\tilde{F}(\bk)|^2\mathcal{K}(\bk)\,,
\end{align}
This follows straightforwardly from the definition of Fourier transform and its translation property and is consistent with the expression found in \cite{pozas2015harvesting}.

Finally, we remark that the usual dipole coupling in light-matter interaction allows for complex-valued smearing functions, e.g. when one considers a hydrogen atom coupled to electric field. So long as there is no exchange of angular momentum involved between the detectors and the field, the results obtained using real-valued smearing and switching functions will be qualitatively similar \cite{pozas2016entanglement}.

\section{Commutator in arbitrary dimensions and strong Huygens' principle}
\label{appendix: strong-Huygens}

In this section we calculate the expression for the (vacuum expectation value of the) field commutator \mbox{$C^-(\sx,\sx')=[\phi(\sx),\phi(\sx')]$} in arbitrary dimensions. We note that this expectation value is state-independent, and all the state-dependence of the Wightman function is contained in the expectation value of the anti-commutator $C^+(\sx,\sx')$. Using the fact that $C^-(\sx,\sx') = W(\sx,\sx')-W(\sx',\sx)$ and Eq.~\eqref{eq: angular-integral} we have
\begin{align}
    C^-(\sx,\sx') &= \frac{\ii}{(2\pi)^n}\frac{2\pi^\frac{n-1}{2}}{\Gamma(\frac{n-1}{2})}\int_0^\infty \dd|\bk|\,|\bk|^{n-2}\int^\pi_0\dd\theta \sin^{n-2}\theta \sin\left(-\ii|\bk|\Delta t + \ii|\bk||\Delta\bx|\cos\theta\right)\,,
\end{align}
where we have used the shorthand $\Delta t = t-t',\Delta x = \bx-\bx'$. Writing $\omega = |\bk|$ and performing the angular integral, we get
\begin{align}
    C^-(\sx,\sx') &= -\frac{\ii}{\sqrt{(4\pi)^n}} 
    \left[2\int_0^\infty \dd\omega\,\omega^{n-2} \sin ( \omega \Delta t ) \, _0\tilde{F}_1\left(\frac{n}{2};-\frac{1}{4} |\Delta\bx|^2 \omega ^2\right)\right]\,,
    \label{eq: commutator-Ndim}
\end{align}
where ${_0\tilde{F}_1}$ is the regularized generalized hypergeometric function \cite{NIST:DLMF}. Note that the term in the square bracket is in the form of Fourier sine transform. As an example, we can readily recover the case for $n=1$, $n=2$ and $n=3$ previously calculated, for example, in \cite{Causality2015Eduardo}:
\begin{align}
    C^-_1(\sx,\sx') &= -\frac{\ii}{\sqrt{4\pi}}\left[2\int_0^\infty\dd\omega\,\sin(\omega\Delta t)\frac{\cos(\omega|\Delta x|)}{\omega\sqrt{\pi}}\right] = -\frac{\ii\,\text{sgn}(\Delta t)}{2}\Theta(|\Delta t| - |\Delta x|)\,,\\
    C^-_2(\sx,\sx') &= -\frac{\ii}{4\pi}\left[2\int_0^\infty\dd\omega\,\sin(\omega\Delta t) J_0(\omega|\Delta \bx|)\right] = -\frac{\ii\,\text{sgn}(\Delta t)}{2\pi}\frac{\Theta(\Delta t^2 - |\Delta \bx|^2)}{\sqrt{\Delta t^2-|\Delta \bx|^2}}\,,\\
    C^-_3(\sx,\sx') &= -\frac{\ii}{(4\pi)^{3/2}}\left[2\int_0^\infty \dd\omega\,\omega \sin(\omega\Delta t)\frac{2\sin(\omega|\Delta \bx|)}{\sqrt{\pi}|\Delta \bx|}\right] = \frac{\ii}{4\pi |\Delta \bx|}\left[\delta(\Delta t + |\Delta\bx|) - \delta(\Delta t - |\Delta\bx|)\right]\,.
\end{align}
The expressions for arbitrary dimensions can be worked out analogously.

Let us consider what happens for the commutator when $n$ is odd and $n\geq 3$. The  strong Huygens' principle says that for odd number of spatial dimensions (odd $n$), the support of the commutator is only along the null direction $\Delta t = \pm |\Delta \bx|$. We will calculate this explicitly for $n=5$ and $n=7$ and provide the generic form for arbitrary odd $n$. The crucial part of the upcoming calculation is that for odd $n\geq 5$, the support is confined to be along the null direction. However, notice that it involves not only Dirac delta functions but also their {distributional derivatives}. Let us denote the distributional derivatives of the Dirac delta function by $\delta^{(k)}(z)$ where $(k)$ denotes the number of derivatives. The distributional derivative has the property that
\begin{align}
    \int_{-\infty}^\infty \dd z\, f(z)\delta^{(k)}(z-z_0) &= (-1)^{k}\frac{\dd^k f}{\dd z^k}(z_0)
    \label{eq: dist-derivatives}
\end{align}
and in particular $\int \dd z\, \delta^{(k)}(z)=0$ for all $k\geq 1$.

In order to calculate the commutator for $n=5$, first we rewrite Eq.~\eqref{eq: commutator-Ndim} as 
\begin{align}
     C^-_5(\sx,\sx') &= -\frac{2\ii}{\sqrt{(4\pi)^5}} \int_0^\infty \dd\omega\, \omega^3\sin(\omega\Delta t) _0\tilde{F}_1\left(\frac{5}{2};-\frac{1}{4} |\Delta\bx|^2 \omega ^2\right) \notag\\
     &= \frac{2\ii}{\sqrt{(4\pi)^5}} \int_0^\infty \dd\omega \frac{2}{\sqrt{\pi}|\Delta\bx|^3}\bigg[\cos(\omega \Delta t+\omega|\Delta\bx|) - \cos(\omega \Delta t-\omega |\Delta\bx|)\bigg] \notag\\
    &+\frac{2\ii}{\sqrt{(4\pi)^5}} \int_0^\infty \dd\omega\, \frac{2\omega}{\sqrt{\pi}|\Delta\bx|^2}
    \bigg[\sin(\omega \Delta t+\omega|\Delta\bx|) + \sin(\omega \Delta t-\omega |\Delta\bx|)\bigg]\,.
\end{align}
Integrating over $\omega$ from $0$ to $\infty$, the first line in the last step is essentially the Fourier cosine transform of a constant function, while the second line is proportional to the Fourier sine transform of $\omega$. Therefore, we obtain
\begin{align}
    C^-_5(\sx,\sx') &= \frac{\ii}{8\pi^2|\Delta\bx|^3}\bigg[\delta(\Delta t+|\Delta \bx|)-\delta(\Delta t-|\Delta \bx|)\bigg] 
    -\frac{\ii}{8\pi^2|\Delta\bx|^2}\bigg[
    \delta^{(1)}(\Delta t+|\Delta \bx|)
    +\delta^{(1)}(\Delta t-|\Delta \bx|)\bigg]\,.
\end{align}
Note that the commutator is supported only along the null direction, but there is a contribution due to first derivative of the Dirac delta function $\delta^{(1)}(\Delta t\pm |\Delta \bx|)$ which dominates for larger $|\Delta\bx|$.

In order to calculate the commutator for $n=7$, first we rewrite Eq.~\eqref{eq: commutator-Ndim} as 
\begin{align}
    C^-_7(\sx,\sx') &= -\frac{2\ii}{\sqrt{(4\pi)^7}} \int_0^\infty \dd\omega\, \omega^5\sin(\omega\Delta t) _0\tilde{F}_1\left(\frac{7}{2};-\frac{1}{4} |\Delta\bx|^2 \omega ^2\right) \notag\\
    &= \frac{2\ii}{\sqrt{(4\pi)^7}} \int_0^\infty \dd\omega\,\frac{12}{\sqrt{\pi}|\Delta\bx|^5}\bigg[\cos(\omega \Delta t+\omega|\Delta\bx|) - \cos(\omega \Delta t-\omega |\Delta\bx|)\bigg] \notag\\
    &+\frac{2\ii}{\sqrt{(4\pi)^7}} \int_0^\infty \dd\omega\, \frac{12\omega}{\sqrt{\pi}|\Delta\bx|^4}
    \bigg[\sin(\omega \Delta t+\omega|\Delta\bx|) + \sin(\omega \Delta t-\omega |\Delta\bx|)\bigg]\notag\\
    &-\frac{2\ii}{\sqrt{(4\pi)^7}} \int_0^\infty \dd\omega\, \frac{4\omega^2}{\sqrt{\pi}|\Delta\bx|^3}
    \bigg[\cos(\omega \Delta t+\omega|\Delta\bx|) - \cos(\omega \Delta t-\omega |\Delta\bx|)\bigg]\,.
\end{align}
Integrating over $\omega$ from $0$ to $\infty$, the first line is the Fourier cosine transform of a constant function, the second line is the Fourier sine transform of $\omega$, and now we also have the third line proportional to Fourier cosine transform of $\omega^2$. Therefore, we obtain
\begin{align}
    C^-_7(\sx,\sx') &= \frac{3\ii}{16\pi^3|\Delta\bx|^5}\bigg[\delta(\Delta t+|\Delta \bx|)-\delta(\Delta t-|\Delta \bx|)\bigg] 
    -\frac{3\ii}{16\pi^3|\Delta\bx|^4}\bigg[
    \delta^{(1)}(\Delta t+|\Delta \bx|)
    +\delta^{(1)}(\Delta t-|\Delta \bx|)\bigg]\notag\\
    & + \frac{\ii}{16\pi^3|\Delta\bx|^3}\bigg[\delta^{(2)}(\Delta t+|\Delta \bx|)-\delta^{(2)}(\Delta t-|\Delta \bx|)\bigg]\,.
\end{align}
Note that the commutator is supported only at the light cone, but there are contributions due to first and second derivatives of the Dirac delta function, with the second derivatives dominating at small distances. 

More generally, following the same procedure one can show that odd $n\geq 3$, the commutator takes the generic form
\begin{align}
    C^-_{n}(\sx,\sx') &= \ii\sum_{j=0}^{\frac{n-3}{2}}\frac{a_j}{|\Delta\bx|^{n-2-j}}\left[\delta^{(j)}(\Delta t+|\Delta \bx|) +(-1)^{j+1}\delta^{(j)}(\Delta t-|\Delta \bx|)\right]\,,
    \label{eq: general-commutator}
\end{align}
where $a_j$ are real constants, and at small distances the commutator is dominated by the highest derivative of the Dirac delta function. Thus we showed that the strong Huygens' principle is satisfied in Minkowski spacetimes with odd number of spatial dimensions $n\geq 3$.

Finally, we remark that the distributional derivatives of the Dirac delta function are responsible for the increasing number of peaks in $|\mathcal{M}^-|$ in higher dimension. Roughly speaking, this comes from the fact that  $\mathcal{M}^-$ is an integral with respect to $t,t'$ over the switching functions multiplied with the commutator $C^-_n(\sx,\sx')$. Since we considered detectors that are sufficiently separated spatially (large enough $|\Delta \bx|$), the dominant contribution comes from the highest ($\frac{n-3}{2}$-th) derivative of the delta function. Therefore, the dominant feature of $|\mathcal{M}^-|$ comes from convolution of the switching functions with the highest derivative, so the peaks of $|\mathcal{M}^-|$ come from the behaviour of the derivatives of the switching functions centred about the null direction.

\section{Wightman function for massive scalar fields in arbitrary spacetime dimensions}
\label{appendix: Wightman-arbitrary}

Here we study the behaviour of the Wightman functions of arbitrary $m\geq 0$ and $n\geq 2$. For completeness we will first derive derive the Wightman function for a massive scalar field in arbitrary dimensions. The Wightman function was also derived in \cite{Takagi1986noise} but the steps in there required restrictions to timelike-separated points. Here we present a more general expression.

Following the procedure in Appendix~\ref{appendix: M-proof}, we know that the Wightman function can be written as
\begin{align}
    W(\sx,\sx') &= \frac{1}{2(2\pi)^n}\frac{2 \pi ^{\frac{n-1}{2}}}{\Gamma \left(\frac{n-1}{2}\right)}\int_0^\infty\dd|\bk| {\frac{|\bk|^{n-1}}{\sqrt{|\bk|^2+m^2}}}e^{-\ii\sqrt{|\bk|^2+m^2}\Delta t} \int_0^\pi\dd\theta \sin^{n-2}\theta\,e^{\ii\bk\cdot \Delta \bx}\,.
\end{align}
where $\Delta t = t-t'$ and $\Delta \bx=\bx-\bx'$. The angular part has been solved in \eqref{eq: angular-integral-final}, but it will be convenient for us to use the Bessel-Clifford functions \eqref{eq: bessel-angle} and write this as
\begin{align}
        W(\sx,\sx') &= \frac{2 \pi ^{\frac{n}{2}}}{2(2\pi)^n}\int_0^\infty\dd|\bk| {\frac{|\bk|^{n-1}}{\sqrt{|\bk|^2+m^2}}}e^{-\ii\sqrt{|\bk|^2+m^2}\Delta t} \left(\frac{2}{|\bk||\Delta\bx|}\right)^{\frac{n-2}{2}}J_{\frac{n-2}{2}}(|\bk||\Delta\bx|)\,.
\end{align}
We perform the change of variable $s=  \sqrt{|\bk|^2+ m^2}/m$, so that
\begin{align}
    W(\sx,\sx') &= \frac{m ^{\frac{n}{2}}}{2(2\pi)^{\frac{n}{2}}|\Delta\bx|^{\frac{n-2}{2}}}
    \int_1^\infty\dd s (s^2-1)^{\frac{n-2}{2}} e^{-\ii m s \Delta t}J_{\frac{n-2}{2}} \left(m|\Delta\bx|\sqrt{s^2-1}\right)\,.
\end{align}
We can now use the identity \#6.645 in  \cite{gradshteyn2014table}:
\begin{align}
    \int_1^\infty\dd x (x^2-1)^{\frac{\nu}{2}} e^{-\alpha x}J_{\nu} \left(\beta\sqrt{x^2-1}\right)=
    \sqrt{\frac{2}{\pi}}\beta^\nu(\alpha^2+\beta^2)^{-\frac{\nu}{2}-\frac{1}{4}}K_{\nu+\frac{1}{2}}(\sqrt{\alpha^2+\beta^2})\,,
\end{align}
where $K_\mu$ is the modified Bessel function of the first kind. Setting $\nu=\frac{n-2}{2}$, $\beta = m|\Delta\bx|$ and analytic continuing using {$\alpha = \epsilon + \ii m \Delta t$} gives
\begin{align}
    W(\sx,\sx') &= \frac{m ^{\frac{n-1}{2}}}{(2\pi)^{\frac{n+1}{2}}}\frac{1}{[-({\Delta t-\ii\epsilon })^2+|\Delta\bx|^2]^{\frac{n-1}{4}}}K_{\frac{n-1}{2}}(m\sqrt{-({\Delta t-\ii\epsilon})^2+|\Delta\bx|^2}) \,,
    \label{eq: Wightman-massive-D-dim}
\end{align}
where this expression should be understood as a (bi-)distribution. The corresponding commutator and anti-commutator can then be obtained using $C^\pm(\sx,\sx') = W(\sx,\sx') \pm W(\sx',\sx)$. Note that the small mass limit $m\to 0^+$ will give us the Wightman function for the massless scalar field in Eq.~\eqref{eq: Wightman-massless-D-dim-main}.

\twocolumngrid

\bibliography{Ref-comm}

\end{document}